\documentclass[graphics, twocolumn, usenatbib]{mn2e}
\usepackage{times}
\usepackage{natbib} 
\usepackage{epsfig} 
\usepackage{graphicx} 
\usepackage{color}
\usepackage{aas_macros} 
\usepackage{amssymb}
\usepackage{amsmath}
\usepackage[title]{appendix}
\usepackage{hyperref}	% Hyperlinks
\hypersetup{colorlinks=true,linkcolor=blue,citecolor=blue,filecolor=blue,urlcolor=blue}

\newcommand{\enzo}{\texttt{Enzo~}}
\newcommand{\grackle}{\texttt{Grackle-2.1~}}
\newcommand{\gracklec}{\texttt{Grackle-2.1}}

\newcommand{\mpch} {\rm $h^{-1}$ Mpc\,\,} 
\newcommand{\kpch} {\rm $h^{-1}$ kpc\,\,} 
\newcommand{\msolar} {$\rm{M_{\odot}}~$}
\newcommand{\msolarc} {$\rm{M_{\odot}}$}
\newcommand{\zsolar} {$\rm{Z_{\odot}}~$}

\newcommand{\molH} {$\rm{H_2}$~}
\newcommand{\molHc} {$\rm{H_2}$}
\newcommand{\J} {$\rm{10^{-21}\ erg\ cm^{-2}\ s^{-1}\ Hz^{-1}\ sr^{-1}}$}

\def\etal{{\it et al.}~}
\begin{document}
\title[]{Positive or Negative? The Impact of X-ray Feedback on the Formation of Direct Collapse Black Hole Seeds}

\author[J.A. Regan \etal] 
{John A. Regan$^{1}$\thanks{E-mail:john.a.regan@durham.ac.uk}, Peter H. Johansson$^{2}$ 
\& John H. Wise$^{3}$ \\ \\
$^1$Institute for Computational Cosmology, Durham University, South Road, Durham, UK, DH1 3LE \\
$^2$Department of Physics, University of Helsinki, Gustaf H\"allstr\"omin katu 2a,
FI-00014 Helsinki, Finland \\
$^3$Center for Relativistic Astrophysics, Georgia Institute of Technology, 837 State Street, 
Atlanta, GA 30332, USA
\\}

%Start things off
%\date{Accepted XXX. Received YYY; in original form ZZZ}
\pubyear{2016}
\label{firstpage}
\pagerange{\pageref{firstpage}--\pageref{lastpage}}

%Make the Title
\maketitle

 %%%%%%%%%%%%%%%%%%%%%%%%%%%%%%%%%%%%%%%%%%%%%%%%%%%
%Abstract time
\begin{abstract} 
A nearby source of Lyman-Werner (LW) photons is thought to be a central component in dissociating \molH 
and allowing for the formation of a direct collapse black hole seed. Nearby sources are also
expected to produce copious amounts of hydrogen ionising photons and X-ray photons. We study here the feedback 
effects of the X-ray photons by including a spectrum due to high-mass X-ray binaries on top of a galaxy with a 
stellar spectrum. We explicitly trace photon packages emerging from the nearby source and track 
the radiative and chemical effects of the multi-frequency source $(E_{\rm photon} = \rm{0.76\ eV \rightarrow 
7500\ eV}$). We find that X-rays have a strongly negative feedback effect, compared to a stellar only source, 
when the radiative source is placed at a separation greater than $\gtrsim 1 \ \rm kpc$. The X-rays heat the low 
and medium density gas in the envelope surrounding the collapsing halo suppressing the mass inflow. The result 
is a smaller enclosed mass compared to the stellar only case. However, for 
separations of $\lesssim 1 \ \rm kpc$, the feedback effects of the X-rays becomes somewhat neutral. The enhanced 
LW intensity at close separations dissociates more \molH and this gas is heated due to stellar photons alone, 
the addition of X-rays is then not significant. This distance dependence of X-ray feedback suggests that a 
Goldilocks zone exists close to a forming galaxy where X-ray photons have a much smaller negative feedback 
effect and ideal conditions exist for creating massive black hole seeds. 
\end{abstract}

%%%%%%%%%%%%%%%%%%%%%%%%%%%%%%%%%%%%%%%%%%%%%%%%%%%
%keywords time
%%%%%%%%%%%%%%%%%%%%%%%%%%%%%%%%%%%%%%%%%%%%%%%%%%%
%Introduction time

%%%%%%%%%%%%%%%%%%%%%%%%%%%%%%%%%%%%%%%%%%%%%%%%%%%%%%%%%%%%%%%%%%%%%%%%%%%%

\section{Introduction} \label{Sec:Introduction}
The discovery of a large number of super-massive black holes (SMBHs) in the early Universe presents a
challenge to our understanding of the formation of compact objects in the first billion years. How could 
such massive objects form and grow to such huge masses so quickly? The most distant SMBH that has been 
observed has a redshift of $z = 7.085$ and a mass of $\sim 2 \times 10^9$ \msolar \citep{Mortlock_2011} while 
the most massive SMBH observed in the early Universe has a mass of $\sim 1.2 \times 10^{10}$ \msolar at a 
redshift of $z = 6.30$ \citep{Wu_2015}. If, as expected, a massive star must be the progenitor for these SMBHs 
then the stellar remnant must grow at enormous rates (most likely at or above the Eddington rate for its 
entire growth phase) to reach the huge black hole masses observed. Simulations of the formation and evolution 
of the first stars show that the characteristic mass of the first metal-free stars is expected to be around 
40 \msolar  \citep{Stacy_2010, Greif_2011, Clark_2011, Bromm_2013, Hirano_2014, Safranek-Shrader_2016, 
Valiante_2016} leading to remnant black hole masses which must grow by up to eight orders of magnitude by 
$z\sim 7$. Further exacerbating the situation is that these Population III (Pop III) stars are expected to 
form in low mass halos (see e.g. \citealt{Bromm_2011}). The resultant supernova are then expected to expel 
the gas from the halo further hampering the growth \citep{Johnson_2007, Milosavljevic_2009, Alvarez_2009, 
Hosokawa_2011} of the black hole and almost certainly restricting the black hole growth to values much less 
than the Eddington rate. All of these obstacles combine to make Pop III stars rather unattractive progenitors 
for the SMBHs observed at early times. \\
\indent If instead we form super-massive stars (SMS), with initial masses of $\gtrsim 10^3$ \msolarc, 
in more massive halos, in the early Universe we can conveniently side-step the growth requirements. The 
initial star grows to super-massive scales via mass accretion (e.g. \citealt{Hosokawa_2013}) reaching a mass of a 
few times $10^5$ \msolar before undergoing a general relativistic instability (e.g. \citealt{Shibata_2016}).
SMS are expected to directly collapse into black holes with masses close to that of the progenitor (see e.g. 
\citealt{Chandrasekhar_1964b}). As a result the black hole gets a head start compared to a comparatively small 
Pop III star. Direct collapse black holes (DCBHs) then offer a promising mechanism to explain the 
existence of quasars at redshifts greater than six. Numerous analytical, semi-analytical and numerical 
studies have been undertaken in recent years to study in great detail the direct collapse mechanism 
\citep{Bromm_2003, Wise_2008a, Regan_2009b, Regan_2009, Tseliakhovich_2010,
Inayoshi_2012, Agarwal_2013, Latif_2013c, Tanaka_2014, Agarwal_2014b, Mayer_2014, Regan_2014a, Regan_2014b, 
Inayoshi_2015}. In order to form a SMS we need to disrupt the usual mechanisms
that lead to the formation of Pop III stars. \molH is the dominant coolant in the early Universe, if this cooling
channel is blocked then the gas will remain at the atomic cooling threshold of $ T \sim 8000$ K assuming 
it is also metal free (for atomic cooling halos with $T_{\rm vir} \sim 10^4$ K). 
Eliminating \molH can be achieved either through photo-dissociation or collisional dissociation. \\
\indent Collisional dissociation of \molH (\molH + H $\rightarrow$ 3 H) is effective for gas 
of a primordial composition and high temperature satisfying the criteria of the ``zone of no-return'' 
\citep{Visbal_2014}. \cite{Inayoshi_2012} suggested that cold accretion shocks may provide a 
pathway to collisionally dissociate \molH during gravitational collapse. However, 
\cite{Fernandez_2014} demonstrated, through numerical simulations, that in the absence of a 
photo-dissociating background this method is difficult to achieve in practice as the 
collisional processes tend to operate at the virial radius and not in the centre of the halo. \\
\indent Photo-dissociation of \molH has been studied by several authors as a viable means 
of disrupting \molH cooling at high redshift where metal cooling is unavailable 
\citep{Omukai_2001, Oh_2002, Bromm_2003, Shang_2010, Latif_2014a, Latif_2014b, Latif_2015}. 
In this case radiation in the Lyman-Werner (LW) band with energies between 11.2 and 13.6 eV 
is able to dissociate \molH via the two step Solomon process \citep{Field_1966, Stecher_1967}. 
\begin{align}
\mathrm{H_2 + \gamma  \rightarrow H_2^{*}} \\
\mathrm{H_2^{*}      \rightarrow H + H + \gamma}
\end{align}
\noindent In order for a halo to receive a large \molH dissociating flux it must be near a luminous star-forming 
galaxy which will irradiate the protogalactic cloud and which may augment an already existing background flux. 
However, star-forming galaxies will also produce copious amount of hydrogen ionising radiation 
(hereafter ionising radiation) which will photo-ionise and heat the gas as well as destroying \molHc. While 
the mean free path of ionising radiation will be much shorter than LW radiation, for halos which are sufficiently 
close the HII region created by the ionising flux will be important. Further study has been dedicated to the study
of X-ray backgrounds which are expected to become relevant as the number density of X-ray sources increases. 
Recently, \cite{Hummel_2015} have investigated the impact of a cosmic X-ray background on Pop III formation 
while both \cite{Inayoshi_2011,Inayoshi_2015b} and \cite{Latif_2015} have investigated the impact of X-ray 
backgrounds on the DCBH paradigm. As these works are closely related to the study here we will reflect on all 
of these studies in \S \ref{Sec:HardXRays} and \S \ref{Sec:Discussion}.  \\
\indent In \cite{Regan_2016a} (R16) we investigated the impact of radiation from a nearby source with photon 
energies up to 60 eV (i.e. stellar only model). We found that for very closely separated halos 
(R $\lesssim 0.5$ kpc) the proto-halo was photo-evaporated while for halos that are too distant 
(R $> 4.0$ kpc) the impact of the LW flux was insignificant. We determined that for halos separated by 
approximately 1 kpc, the flux received from a single nearby realistic galaxy resulted in the formation of a 
large core\footnote{The core of the halo is defined 
at the point where the baryonic mass exceeds the dark matter mass. This fluctuates between 
approximately 1 pc and 5 pc across the simulations. We therefore choose 1 pc to define the radius 
of the core of the halo in all cases for consistency.} mass of close to $M_{\rm{core}} \sim 10^4 $\msolar 
with a core temperature of $T \sim 1000$ K surrounded by a large reservoir of warm gas (T$_{\rm{vir}} \sim 10^4$ K).
Such an environment should represent an ideal location for forming a SMS. \\
\indent In this paper we expand on our previous study by also considering the 
impact of both soft and hard X-rays. Nearby galaxies as well as supplying
a strong source of LW and ionising photons should also produce a supply of X-ray photons through the formation of 
high-mass X-ray binaries (HMXBs) as massive stars reach the end of the lifetimes. \textit{The goal of this 
paper is then to investigate this important scenario and to determine whether X-rays have a negative or 
positive effect on the direct collapse scenario when a collapsing halo is irradiated by an anisotropic source.} 
As in R16 our intention is therefore not to investigate the numerical value of 
``$\rm{J_{crit}}$''\footnote{$\rm{J_{crit}}$ is taken to be value of the background radiation intensity required to 
fully dissociate \molH from a target halo.} in this 
instance but rather taking the results of the ``Renaissance'' Simulation suite (see \S \ref{Sec:RadiationSource}) 
to investigate the impact of a realistic source on a nearby galaxy. Our results, similar to R16, will in fact show 
that achieving complete \molH dissociation through irradiation from a single close-by neighbour is very 
unlikely (see R16 for a comprehensive discussion on this topic) and will require (if full \molH dissociation is 
indeed ever required) more than one nearby source. In this sense we do not simulate the classical 
DCBH formation case and rather we instead focus on simulating realistic environments from first principles 
without invoking idealised conditions (e.g. ultra-strong radiation fields) conductive to DCBH formation.\\
\indent The paper is laid out as follows: in \S \ref{Sec:Model} we describe the 
model setup and the numerical approach used, the chemical model and radiation 
prescription employed; in \S \ref{Sec:Results} we describe the results of our 
numerical simulations; in \S \ref{Sec:Discussion} we discuss the importance of the results 
and in \S \ref{Sec:Conclusions} we present our conclusions.  
Throughout this paper we  assume a standard $\Lambda$CDM cosmology with the following parameters 
\cite[based on the latest Planck data]{Planck_2014}, $\Omega_{\Lambda,0}$  = 0.6817, 
$\Omega_{\rm m,0}$ = 0.3183, $\Omega_{\rm b,0}$ = 0.0463, $\sigma_8$ = 0.8347 and $h$ = 0.6704. 
We further assume a spectral index for the primordial density fluctuations of $n=0.9616$.

%%%%%%%%%%%%%%%%%%%%%%%%%%%%%Table 1%%%%%%%%%%%%%%%%%%%%%%%%%%%%%%%%%%%%%%%%%%%%%%%%%%%%%
\begin{table*}
\centering
\caption{Radiation SED}
%\begin{tabular}{ | l | c | c }
\begin{tabular*}{0.99\textwidth}{@{\extracolsep{\fill}} lcc}
\hline 
\textbf{\em $\rm{Spectrum}$}
& \textbf{\em $\rm{Energy\ Bins\ (eV)}$} & \textbf{\em $\rm{Photon\ Fraction\ (PF)}$} \\
\hline 

Stellar                  & 0.76,   8.0,    12.8,   14.79,  20.46,  27.62, 60.0    & 
                           0.4130, 0.3170, 0.1080, 1.32e-07, 2.23e-04,  3.49e-03,  2.26e-02 \\
Stellar + Soft X-rays     & 0.76,   8.0,    12.8,   14.54,  21.87,  119.67, 380.12 & 
                           0.4130, 0.3170, 0.1080,  6.65e-08, 1.22e-04,  1.78e-02,  9.53e-03 \\
Stellar + Hard X-rays     & 0.76,   8.0,    12.8,   17.84,  25.06,  52.93, 69.47,          &
                           0.4130, 0.3170, 0.1080,  6.21e-06, 4.42e-04,  5.05e-03,  8.29e-03, \\
                         & 137.11, 252.82,  750.29,  7570.53 & 
                          6.923-03, 9.59e-03, 5.77e-03, 7.05e-03\\

\hline
\end{tabular*}
\parbox[t]{0.99\textwidth}{\textit{Notes:} The energy bins and the fractional number of photons 
are given for the stellar spectrum and the stellar + X-ray spectrum for the cases of both soft $(< 1\ \rm keV)$ 
and hard X-rays $(> 1\ \rm keV)$. The photon fractions are given for all three cases. In each case the 
photon energies and fractions are identical for energies below the ionisation threshold of hydrogen. For 
energies above the ionisation threshold the sampling energies and sampling fractions are taken from the  
\texttt{sedop} code developed by \cite{Mirocha_2012} which optimises the number and position of the 
energy bins required. 
}

\label{Table:radiation_sed}
\end{table*}
%%%%%%%%%%%%%%%%%%%%%%%%%%%%%%%%%%%%%%%%%%%%%%%%%%%%%%%%%%%%%%%%%%%%%%%%%%%%%%%%%%%%%%%%%%%%%%%%
\section{Model Setup} \label{Sec:Model}
The numerical model used in this study is very similar to the model used in R16. The 
significant difference is that in this work the effect of X-ray radiation is included 
in the model. Furthermore, compared to R16 an additional realisation is used.
We refer to the first halo as Halo A (this is the same halo as used in R16) and the 
second halo as Halo B. 

\subsection{Numerical Framework} \label{Sec:Framework}
\noindent We ran our simulations using the publicly available adaptive mesh refinement
(AMR) code \texttt{Enzo~}\citep{Enzo_2014}\footnote{http://enzo-project.org/}. In particular we use
version 3.0\footnote{Changeset: 7f49adb4c9b4} which is
the bleeding edge version of the code incorporating a range of new features. We created a fork
off the 3.0 mainline and included improved support for radiative transfer based on the Moray 
implementation of \cite{WiseAbel_2011} and chemical modelling using the \texttt{Grackle} library.
For a more in depth discussion of the ray tracing elements and of the modifications to the chemical
network see R16. \\
\indent All simulations are run within a box of 2 \mpch (comoving), the root grid size is $256^3$ and 
we employ three levels of nested grids. The grid nesting and initial conditions are created using 
the MUSIC software package \citep{Hahn_2011}. Within the most refined region (i.e. level 3) the dark 
matter particle mass 
is $\sim$ 103 \msolarc. In order to increase further the dark matter resolution of our simulations 
we split the dark matter particles according to the prescription of \cite{Kitsionas_2002} and 
as described in \cite{Regan_2015}. We split particles centered on the position of the final collapse
as found from lower resolution simulations within a region with a comoving side length of 43.75 h$^{-1}$ kpc.
Each particle is split into 13 daughter particles resulting in a final dark matter particle mass of $\sim$ 
8 \msolar in the high resolution region. The particle splitting is performed at a redshift of $z=40$ well
before the collapse of the target halo. Convergence testing to study 
the impact of lower dark matter particle masses on the physical results was conducted as discussed in R16. 
All of the simulations are started from a redshift of $z = 100$.\\
\indent The baryon resolution is set by the size of the grid cells, in the highest resolution region 
this corresponds to approximately 0.48  \kpch comoving (before adaptive refinement).  The  maximum 
refinement level for all of the simulations was set to 16 leading to a maximum spatial resolution of 
$\Delta x \sim 5 \times 10^{-3}$ pc at a redshift of $z = 25$. The refinement criteria used in this work 
were based on three physical measurements: (1) The dark matter particle over-density, 
(2) The baryon over-density and (3) the Jeans length. The first two criteria introduce additional 
meshes when the over-density (${\Delta \rho \over \rho_{\rm{mean}}}$) of a grid cell with respect to 
the mean density exceeds 8.0 for baryons and/or DM. Furthermore, we set the 
\emph{MinimumMassForRefinementExponent} parameter to $-0.1$ making the simulation super-Lagrangian 
and therefore reducing the threshold for refinement as higher densities are reached. For the final 
criteria we resolve the local Jeans length by at least 16 cells in these runs. All simulations are run 
until they reach the maximum refinement level at which point the simulation is terminated.\\

\subsection{Radiation Source} \label{Sec:RadiationSource}
As in R16 we use a radiation source to model the impact of a nearby galaxy on a collapsing halo. 
The radiation source is a point particle. It is massless and is fixed in comoving space. 
The physical distance between the source and the collapsing halo therefore inevitably increases 
due to the expansion of the Universe as a function of decreasing redshift. 
The source of radiation is placed at a distance of between 1 kpc and 4 kpc, depending on the given 
model being tested, from the point of maximum density at a redshift of $z=40$. In each case, 
we use a luminosity of $1.2 \times 10^{52}$ photons per second (above the H$^-$ photo-detachment 
energy of 0.76 eV) that originates from a galaxy with a stellar mass of $10^3$ \msolar at $z = 40$. 
The galaxy has a  specific star formation rate (SFR) of $\rm sSFR=40 \ \rm Gyr^{-1}$ resulting in a stellar mass of 
$10^5$ \msolar at $z = 20$.  The stellar mass at $z = 20$ and the specific SFR are 
consistent with the largest galaxies prior to reionisation in the Renaissance Simulations of 
\cite{Chen_2014}.  We then calculate its spectrum with the \cite{Bruzual_2003} models with a 
metallicity of $10^{-2}$ \zsolar and compute the photon luminosity from it. Furthermore, for the 
models which include X-rays we include the contribution from six HMXB sources (see \S
\ref{Sec:XRays}). The spectrum does not include emission from the nebular component and is solely due to 
stellar and X-ray emission from individual sources. \\ 
%%%%%%%%%%%%%%%%%FIGURE 1%%%%%%%%%%%%%%%%%%%%%%%%%%%%%%%%%%%%%%%%%%%%%%%
\begin{figure*}
  \centering 
  \begin{minipage}{175mm}      \begin{center}
      \centerline{
        \includegraphics[width=9cm]{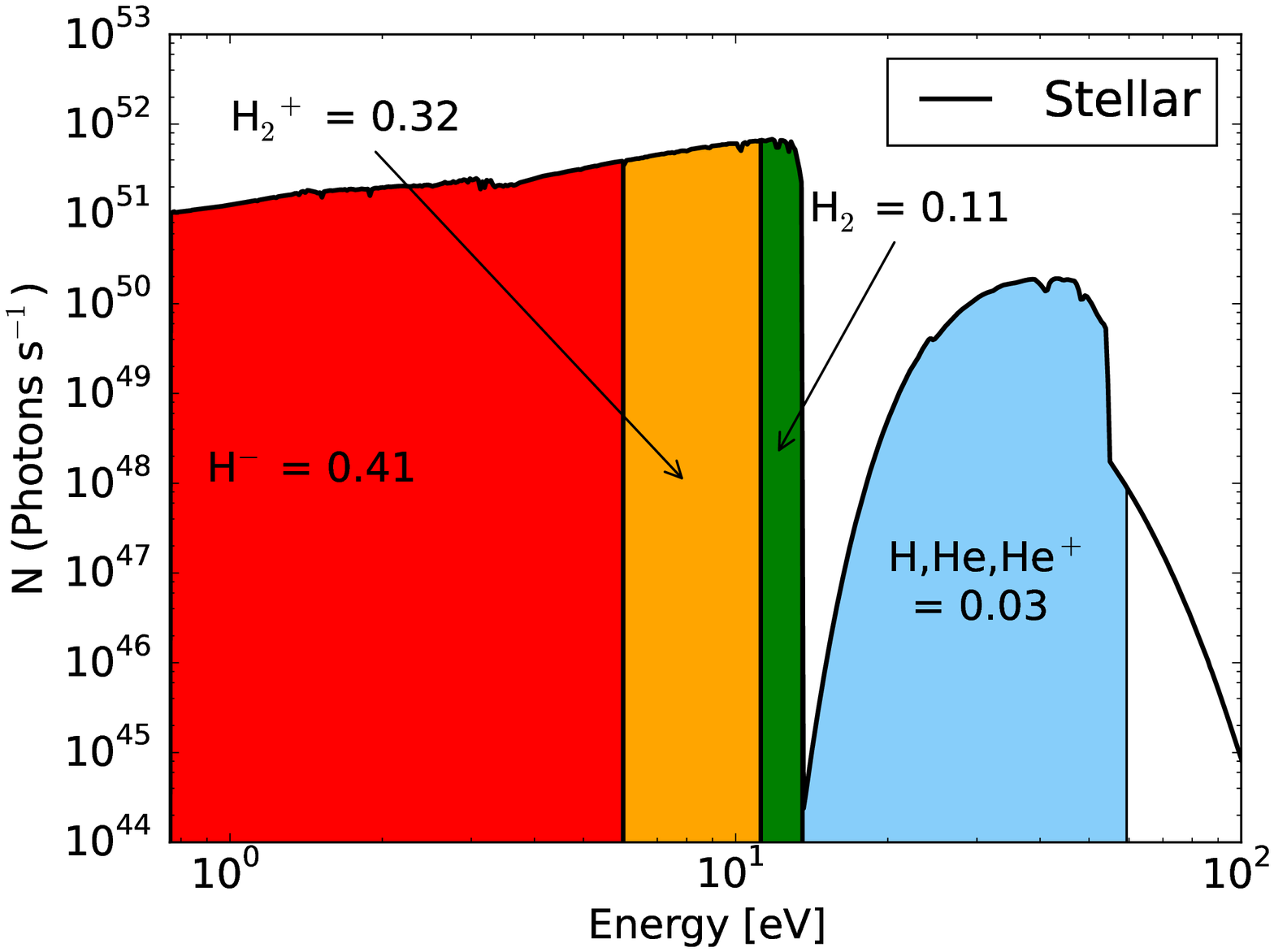}
        \includegraphics[width=9cm]{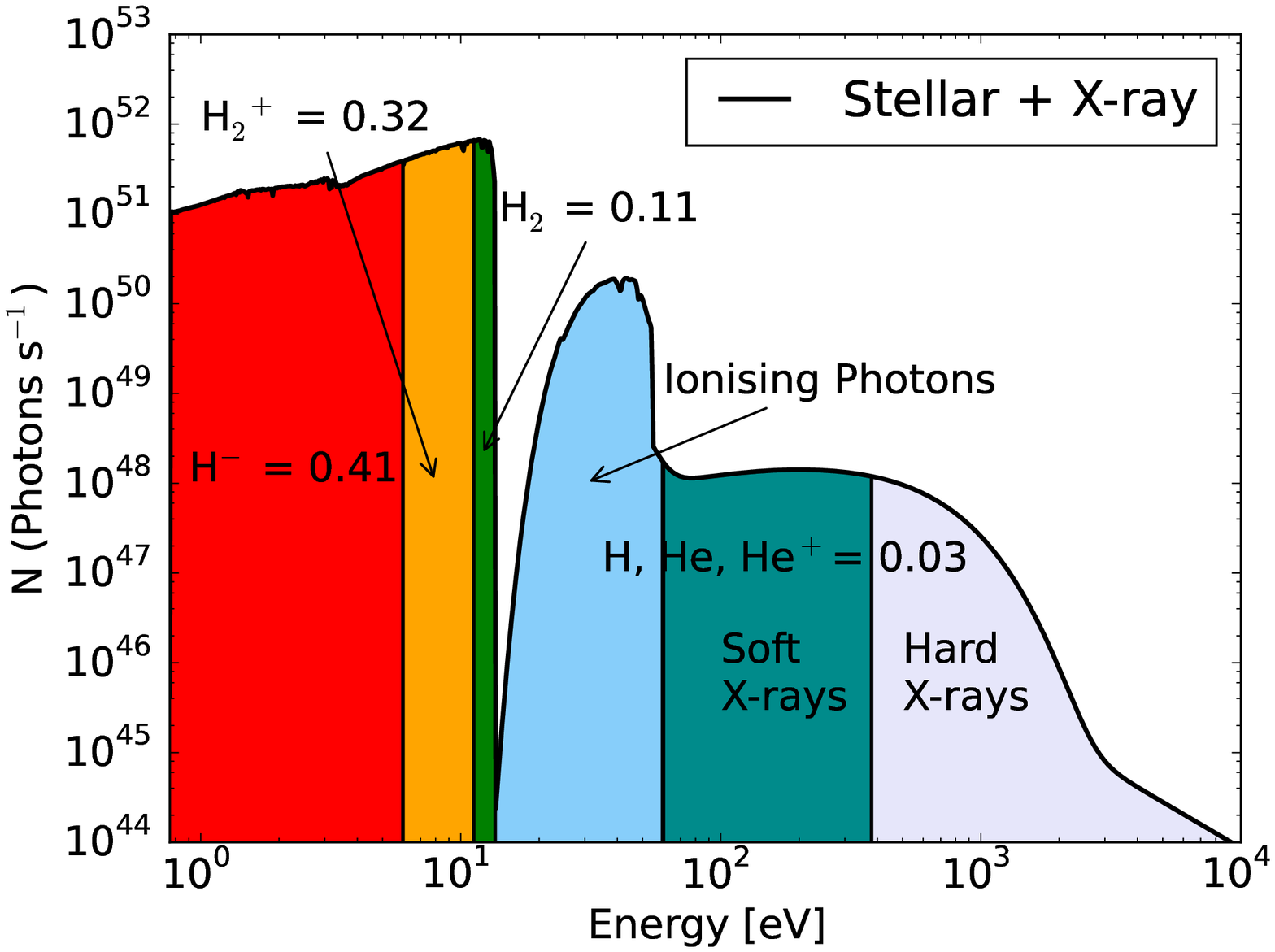}}
        \caption[]
        {\label{StellarSED}
        The left panel shows the luminosity from a stellar spectrum consistent with the
        Renaissance Simulation of \cite{Chen_2014}. The total stellar mass giving rise to this 
        spectrum is $10^5$ \msolar at z = 20. We have employed an extinction 
        factor for photons with energy greater than 13.6 eV and a cutoff for photons 
        greater than 60 eV. In the right hand panel we show the same plot with the inclusion
        of X-rays evenly split between a non-thermal source and a multi-colour disk component.
        The fraction of photons in each energy band is indicated. In both panels the number of 
        photons in each bin is almost the same as the vast majority of the photons have energies less
        than the ionisation threshold for hydrogen. The main difference therefore is that 
        the X-rays are in addition sampled in the right hand plot. The contribution of X-ray photons 
        to the total number photons is relatively small.
        }
      \end{center} \end{minipage}
  \end{figure*}

%%%%%%%%%%%%%%%%%%%%%%%%%%%%%%%%%%%%%%%%%%%%%%%%%%%%%%%%%%%%%%%%%%%%%%%%%%%%

\subsection{Radiation Fields}        \label{Sec:RadiationFields}
In total three different radiation fields were used in this study. The three fields are detailed in Table 
\ref{Table:radiation_sed}. The first field has contributions from a stellar source only. The last two fields
are broken into two parts both with contributions from stellar and X-ray components. The second field in the table
is composed of radiation from a stellar component and a soft X-ray component, with energies up to 380 eV. 
The third field in the table extends the X-ray contribution into the hard X-ray regime with contributions 
of energies up to 7570 eV. \\
\indent The optimal energy bins with which to model our spectra are computed using the  \texttt{sedop} code 
developed by \cite{Mirocha_2012}. The \texttt{sedop} code determines the optimum energy and intensity for a 
given number of energy bins needed to accurately model radiation with energy above the ionisation 
threshold of hydrogen. The energy bins below the ionisation threshold are set to capture the peak of the 
photo-detachment of  $\rm{H^-}$ at 0.76 eV, photo-ionisations of $\rm{H_2^+}$ at 8.0 eV and photo-ionisations of 
$\rm{H_2}$ at 12.8 eV. \\
\indent The shape of the stellar spectrum used in this study (see Figure \ref{StellarSED}) is identical to 
the one used in R16. In particular the left hand panel of Figure \ref{StellarSED} includes only the spectrum 
due to a stellar component and is described in \S \ref{Sec:RadiationSource}.  
The right hand panel shows the extra contribution to the spectrum when X-rays are included. For including the 
X-ray contribution to the spectrum we assume that the X-ray luminosity is evenly split between a
multi-color disk component \citep{Mitsuda_1984} formed from the accretion disk feeding the black hole and a  
non-thermal component (power-law) formed from the comptonisation of electrons, originating in the disk, in the 
hot corona surrounding the black hole.  This model is similar to that used by numerous models of black hole 
spectra in the literature \citep{Zdziarski_2001, Kuhlen_2005, Done_2007}. We assume that there are six 
HMXB sources, as was typically observed in the Renaissance simulations \citep{Xu_2013}, active within the 
galaxy, we take a mass of 40 \msolar for each of the black holes and finally 
we assume a radiative efficiency of 0.1 times Eddington. The photon fraction (i.e. SED component) in each energy 
bin is then taken from the spectrum. \\
\indent Finally, we break the X-ray spectra into two further models. For the first 
model we take into account the contribution of a stellar component and a soft X-ray component and impose a 
cut-off at $\sim$ 380 eV, we refer to this model as the \textbf{soft X-ray model}. For the second X-ray model we 
take both the soft and hard components of the spectrum into account as well as the stellar component and 
allow the X-rays to reach energies up to $\sim$ 7500 eV, we refer to this model as the \textbf{hard X-ray model}.
Each of three models includes a stellar component with energies up to $\sim$ 60 eV. 

\subsection{Modelling absorption due to gas in the Interstellar Medium}
\indent We also model the impact of interstellar absorption of ultra-violet photons (with energies 
greater than 13.6 eV) in our model. The impact of this modelling can be seen in the sharp drop in photon numbers 
above 13.6 eV. The model convolves the spectral energies of our spectra with a simple 
modelling of the optical depth to ionising radiation as follows:

\begin{equation}
\rm{PF_{ext}(E)} = PF(E) \times \rm{exp}(-\sigma(E) \times N(HI)_{avg})
\end{equation}
where $\rm PF(E)$ is the photon fraction at the energy, $\rm E$,  $\rm PF_{\rm{ext}}(E)$ is the 
photon fraction when the extinction is accounted for, $\sigma(\rm E)$ is the cross 
section of hydrogen at that energy and $\rm N(HI)_{\rm{avg}}$ is the column density of
hydrogen averaged over the source galaxy. For our model we choose an average value of 
N(HI)$_{\rm{avg}}$ of $2.5 \times 10^{18} \rm{cm^{-2}}$ consistent with the results from the simulations of 
\cite{Wise_2009}. A full description of the physical motivations of this model along with the assumptions 
incorporated into the model is given in R16.

%Physically this is motivated by the fact that low density channels of 
%neutral hydrogen allow for the escape of ionising radiation between approximately 13.6 and 50 eV
%from the radiating galaxy. These channels are somewhat transient and evolve over time \citep{Wise_2014} meaning 
%that over a sufficient amount of time (approximately 80 Myrs) the halo receiving the flux is 
%swept over by ionising radiation in these bands rather than being illuminated constantly. Our 
%ISM modelling is an attempt to take this effect into account. The extinction factor is set to 1.0 for energy 
%below the ionisation threshold of neutral hydrogen, thus having no effect in that case. The mean free path 
%of photons below the ionisation threshold of hydrogen is comparatively long and is not included in our model.
%Strong internal Lyman-Werner flux will dissociate most of the \molH in the source galaxy with the exception of 
%some molecular clouds, which have a small geometric cross-section and can be safely ignored. Likewise the mean
%free path of photons significantly above the ionisation threshold of  hydrogen is comparatively long and so 
%the ISM has little impact on these photons. Hence for photons with energies greater than $\sim 50$ eV the 
%extinction modelling has no impact. 
%%%%%%%%%%%%%%%%%%%%%%%%%%%%%%%Table 2%%%%%%%%%%%%%%%%%%%%%%%%%%%%%%%%%%%%%%%%%%
\begin{table*}
\centering
\caption{Realisations \& Models}
%\begin{tabular}{ | l | c | c | l | c | c | c | c |}
\begin{tabular*}{0.99\textwidth}{@{\extracolsep{\fill}} lccccccc}
\hline 
\textbf{\em {Sim Name}} &
\textbf{\em {Init. Dist. (kpc)}} & \textbf{\em Spectrum} & \textbf{\em{z$_{\rm{coll}}$}} 
& \textbf{\em{Final Dist. (kpc)}} & \textbf{\em{T$_{\rm{vir}}$ (K)}} & \textbf{\em{M$_{200}$ (\msolarc)}} 
& \textbf{\em{M$_{\rm{core}}$ (\msolarc)}} \\
\hline 
1kpc\_S\_A  & 1.0 & Stellar SED               & 25.25 & 1.9 & 6224 & $1.04 \times 10^7$ & 9476  \\
2kpc\_S\_A  & 2.0 & Stellar SED               & 28.67 & 2.9 & 4225 & $4.84 \times 10^6$ & 7269  \\
4kpc\_S\_A  & 4.0 & Stellar SED               & 29.97 & 5.4 & 3181 & $2.96 \times 10^6$ & 6117  \\
1kpc\_X\_A  & 1.0 & Stellar + XRay SED        & 25.13 & 1.9 & 6513 & $1.12 \times 10^7$ & 8475  \\
2kpc\_X\_A  & 2.0 & Stellar + XRay SED        & 29.06 & 2.9 & 3849 & $4.12 \times 10^6$ & 5903  \\
4kpc\_X\_A  & 4.0 & Stellar + XRay SED        & 31.06 & 5.2 & 2212 & $1.63 \times 10^6$ & 3092  \\
1kpc\_HX\_A  & 1.0 & Stellar + Hard XRay SED  & 24.54 & 2.0 &  8675 &  $1.74 \times 10^7$ & 5174 \\
2kpc\_HX\_A  & 2.0 & Stellar + Hard XRay SED  & 29.48  &  2.8 & 3434 & $3.40 \times 10^6$ & 5692 \\
4kpc\_HX\_A  & 4.0 & Stellar + Hard XRay SED  & 31.08  &  5.2 & 2210 & $1.63 \times 10^6$ & 3040  \\

1kpc\_S\_B  & 1.0 & Stellar SED       & 21.41 & 1.4 & 9830 & $2.62 \times 10^7$ & 7587  \\
2kpc\_S\_B  & 2.0 & Stellar SED       & 28.44 & 2.5 & 4332  & $5.08 \times 10^6$  & 10936 \\
4kpc\_S\_B  & 4.0 & Stellar SED       & 29.97  & 5.1 & 3253 & $3.06 \times 10^6$ & 6562  \\
1kpc\_X\_B  & 1.0 & Stellar + XRay SED       & 21.98 & 1.4 & 11447 & $3.17 \times 10^7$ & 6397 \\
2kpc\_X\_B  & 2.0 & Stellar + XRay SED       & 27.89 & 2.5 & 4883 & $6.26 \times 10^6$ & 7677 \\
4kpc\_X\_B  & 4.0 & Stellar + XRay SED       & 31.19 & 4.9 & 2170  & $1.58 \times 10^6$ & 2882  \\
\hline
\end{tabular*}
\parbox[t]{0.99\textwidth}{\textit{Notes:} Each model is run with the radiation source at an 
initial distance from the centre of the collapsing halo of 1.0, 2.0 and 4 kpc (physical). 
The initial distance is the distance at z = 40. For each of these models the spectrum is varied between 
a stellar SED (maximum photon energy = 60 eV and indicated with a ``\_S'' in the name) and a stellar + XRay 
spectrum (indicated by an ``\_X'' in the name). 
The soft X-ray spectrum has energies up to $\sim$ 380 eV while the models including hard X-rays have energies 
up to $\sim 7500$ eV (the simulations including hard X-rays have an ``\_HX'' in their name). 
All distances are in physical kpc unless explicitly stated. The core mass in the final column denotes the 
baryonic mass inside a 1 pc radius around the densest point. 
}
\label{Table:radiation_particle}
\end{table*}

%%%%%%%%%%%%%%%%%%%%%%%%%%%%%%%%%%%%%%%%%%%%%%%%%%%%%%%%%%%%%%%%%%%%%%%%%%%%%%%%%%%%%%%%%%%%%

\subsection{Modelling the contribution due to X-rays} \label{Sec:XRays}
The major difference between this work and that of R16 is the inclusion of an X-ray component. The ionisation 
cross-sections of neutral hydrogen and helium drop off as $\sigma_H (\nu) \propto \nu^{-3}$ and 
$\sigma_{He} (\nu) \propto \nu^{-2}$, respectively as the photon energy increases. As a result X-ray photons have a 
much longer mean free path than ionising photons with energies close to 13.6 eV. To model the X-ray photon effect
on the gas we make use of the ray-tracing capabilities of \enzo \citep{AbelWandelt_2002, WiseAbel_2011}. 
Within \enzo X-rays are defined as photons with energies greater than 100 eV. As a consequence and based
on the results of \texttt{sedop} we have two X-ray energy bins with a  soft X-ray spectrum and
 four energy bins with a hard X-ray spectrum. For each energy bin, including X-rays, 768 ($12 \times 4^3$; 
Healpix level 3) rays are isotropically cast with the energy associated with that bin. Consequently, the number 
of photons per each initial ray is

\begin{equation}
P_{\rm init} =  {L_{\rm gal} \times \rm{Photon Fraction} \times dt_{\rm ph} \over  768 \times E_{\rm ph}}
\end{equation}

where $\rm{L_{gal}}$ is the total bolometric luminosity of our galactic source ($1.64 \times 10^{41}$ erg/s), 
Photon Fraction is the fraction of the photons in a given energy bin (see Table \ref{Table:radiation_sed} for 
values),  $\rm{dt_{ph}}$ is the photon timestep used and $\rm{E_{ph}}$ is the photon energy for that ray. Each 
ray is traced until most of its photons are absorbed (99.99999\%) or the photon reaches the end of its region 
of influence,  which we set as 10\% of the computational domain. As rays propagate through the computational 
domain they are split based on the HEALPix formalism. \\
\indent As the X-ray photons propagate into the surrounding medium they interact with the gas in two ways: they
(1) ionise the hydrogen and helium and (2) they heat the gas\footnote{Both infrared photons and ionising photons 
also heat the gas but to a much lesser extent (see Figure \ref{RayProfiles_1kpc}).}. Since X-ray photons have energies in 
excess of the double ionisation threshold of helium the X-ray photons can photoionise H, He and He$^+$ with 
the respective photoionisation rates:

\begin{equation}
\begin{split}
  k_{\rm ph,H} & = {P_{in}(1 - e^{-\tau_H})(E_{ph}Y_{k, H}/E_{i,H}) \over n_{H} (\Delta x)^3 dt_{ph}} \\ \\
  k_{\rm ph,He} & = {P_{in}(1 - e^{-\tau_{He}})(E_{ph}Y_{k, He}/E_{i,He}) \over n_{He} (\Delta x)^3 dt_{ph}} \\ \\
   k_{\rm ph,He^+} & = {P_{in}(1 - e^{-\tau_{He^+}}) \over n_{He^+} (\Delta x)^3 dt_{ph}} \\ \\
\end{split}
\end{equation}
 where $\rm{P_{in}}$ is the number of photons entering a cell, $\tau_H = n_{H}\sigma_H(E) dl$ is the optical depth 
in that cell, $n_{H}$ is the hydrogen number density, $\sigma_H(E)$ is the energy dependent hydrogen photoionisation 
cross section \cite{Verner_1996}, $dl$ is the path length through that cell and E$_{i}$ are the ionisation 
thresholds for H, He and He$^+$ respectively. All of the above hydrogen subscripts apply equally to helium and 
ionised helium. The factors Y$_k$ are the energy fractions used for the ionisation when secondary ionisations are
also considered \citep{Shull_1985}. In the case of secondary ionisations the primary electron which is freed in 
the original ionisation is free not only to heat the gas but also to cause further ionisations due to its large
kinetic energy. The secondary ionisation is then more effective than the primary ionisation when considering
X-ray ionisations of H and He. For He$^+$, however, the impact of secondary ionisations are not important 
\citep{Shull_1985}.\\
\indent Finally, photons also heat the gas through both excess energy heating and Compton heating. The excess 
energy above the ionisation threshold for each ion, E$_{i}$, heats each of the ions according to

\begin{equation}
\Gamma_H = { P_{in}(1 - e^{-\tau_H})E_{ph}Y_{\Gamma} \over n_{H} (\Delta x^3) dt_{ph}}
\end{equation}
with the same equation applying equally to the helium ions, where $\Gamma_H$ is the heat imposed on species H,  
$Y_{\Gamma}$ is the fraction of energy deposited as heat when secondary ionisations are taken into account. 
The X-rays can also scatter off and heat an electron leading to an extra contribution of the form

\begin{equation}
\Gamma_C = { P_{in}(1 - e^{-\tau_e})\Delta E(T_e) \over n_{H} (\Delta x)^3 dt_{ph}}
\end{equation}
where $\tau_e = n_e \sigma_{KN} dl$ is the optical depth, $n_e$ is the electron number density, $\sigma_{KN}$ 
is the non-relativistic Klein-Nishina cross-section \citep{Rybicki_1979} and $\Delta E(T_{e}) = 4 k_{b}T_e 
(E_{ph}/m_ec^2)$ is the non-relativistically transferred energy to an electron at $T_{e}$ \citep{Ciotti_2001}. It 
should also be noted that in this case the photon continues to propagate and is not absorbed. As a result the
total heating rate is 

\begin{equation}
\Gamma_{\rm Total} = n_H \Gamma_H + n_{He} \Gamma_{He} + n_{He^+} \Gamma_{He^+} + n_e \Gamma_C
\end{equation}

\noindent A model similar to this was previously implemented and tested in \enzo by \cite{Kim_2011} although in 
their case the energy of the X-rays was fixed at 2 keV and the context was the exploration of the feedback 
from black holes at a much lower redshift of $z\sim 3$. 
%%%%%%%%%%%%%%%%%FIGURE 2%%%%%%%%%%%%%%%%%%%%%%%%%%%%%%%%%%%%%%%%%%%%%%%
\begin{figure*}
  \centering 
  \begin{minipage}{175mm}      \begin{center}
      \centerline{
        \includegraphics[width=16.0cm,height=9.75cm]{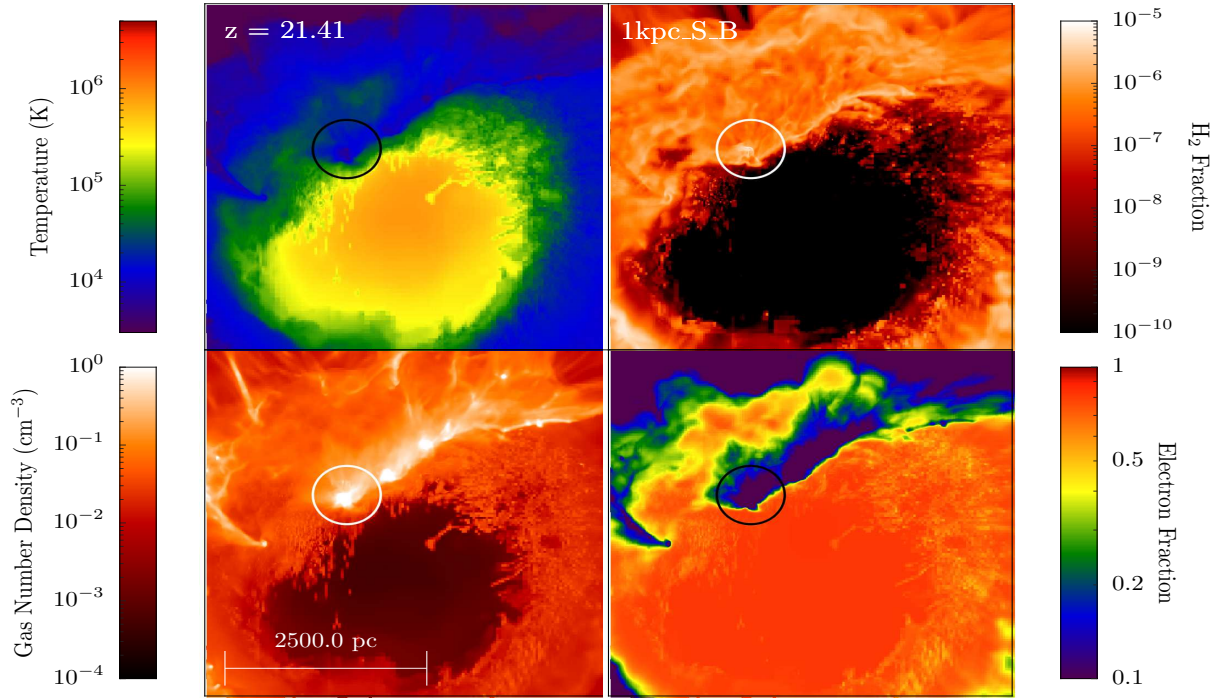}}
        
        \caption[]
        {\label{HaloBProjection_Stellar} HaloB: Stellar SED. Top Left Panel - Temperature, 
          Top Right Panel - \molH Fraction Bottom Left Panel - gas number density, 
          Bottom Right Panel - Electron Fraction. This visualisation is created by projecting
          through a cuboid with dimensions of 2500, 1250, 2500 pc centred on the point of maximum density. 
          The projection is made along the y-axis. The output is the 
          final output time from the 1kpc\_S\_B simulation. The heart shaped region created 
          by the ionising source is clearly visible in each panel. The black or white circle in each panel 
          indicates the position of maximum density, the radius of the circle corresponds to the
          virial radius of the collapsing halo. Each panel is centred on the position of the 
          radiating source at this output time.
        }
      \end{center} \end{minipage}
  \end{figure*}

%%%%%%%%%%%%%%%%%%%%%%%%%%%%%%%%%%%%%%%%%%%%%%%%%%%%%%%%%%%%%%%%%%%%%%%%%%%%

%%%%%%%%%%%%%%%%%FIGURE 3%%%%%%%%%%%%%%%%%%%%%%%%%%%%%%%%%%%%%%%%%%%%%%%
\begin{figure*}
  \centering 
  \begin{minipage}{175mm}      \begin{center}
      \centerline{
        \includegraphics[width=16.0cm,height=9.75cm]{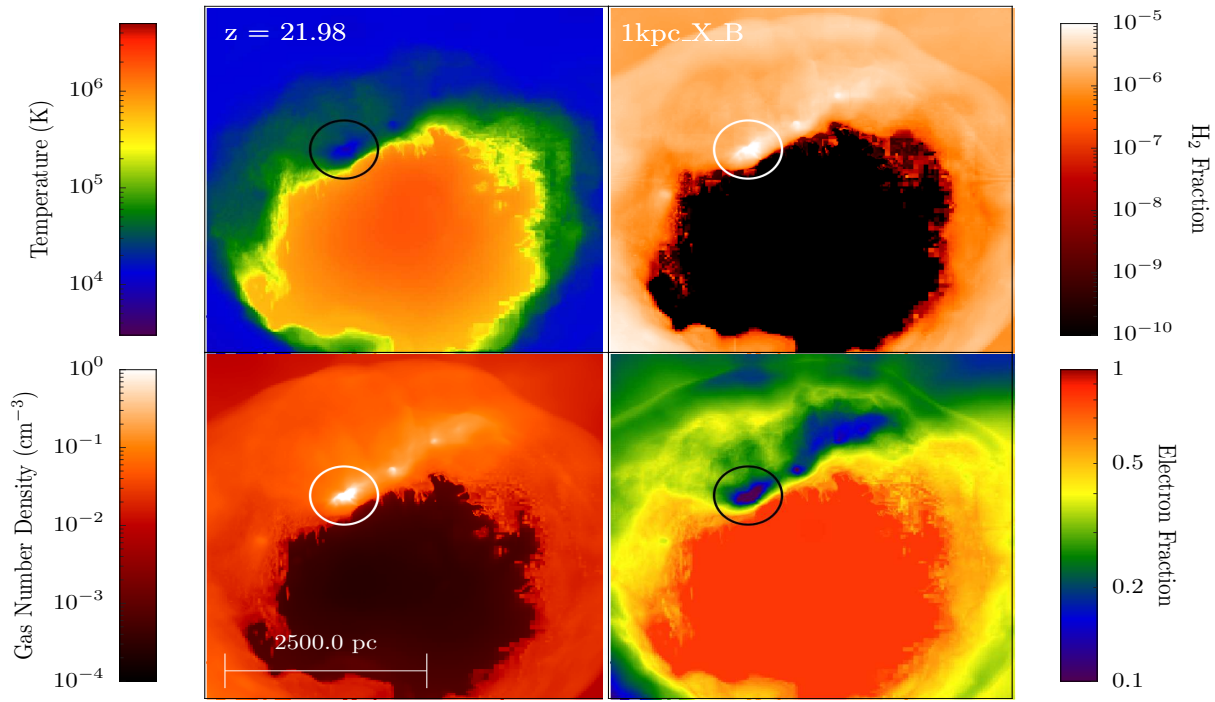}}
        
        \caption[]
        {\label{HaloBProjection_XRays} HaloB: Stellar + XRay SED. Top Left Panel - Temperature, 
          Top Right Panel - \molH Fraction Bottom Left Panel - gas number density, 
          Bottom Right Panel - Electron Fraction. Same as Figure \ref{HaloBProjection_Stellar}
          for simulation 1kpc\_X\_B.
        }
      \end{center} \end{minipage}
  \end{figure*}

%%%%%%%%%%%%%%%%%%%%%%%%%%%%%%%%%%%%%%%%%%%%%%%%%%%%%%%%%%%%%%%%%%%%%%%%%%%%

\subsection{Chemical Network}     \label{Sec:ChemicalNetwork}
We adopt here the 26 reaction network determined by \cite{Glover_2015a} as the most appropriate
network for solving the chemical equations required by the direct collapse model in a gas of 
primordial composition with no metal pollution. The network consists of ten individual species:
${\rm H}, {\rm H}^+, {\rm He}, {\rm He}^+,  {\rm He}^{++}, {\rm e}^-,$ 
$\rm{H_2}, \rm{H_2^+}, \rm{H^-} \rm{and}\ \rm{HeH^+}$. Additionally, we included a further 7 
reactions which accounts for the recombinations (4) and photo-ionisations (3) of  ${\rm H}$, 
${\rm He}$, and ${\rm He}^{+}$ which occurs when the elements are photo-ionised due to photon energies greater 
than 13.6 eV, 25.4 eV and 54.4 eV, respectively. \\
\indent To implement the chemical network we have extensively modified the open source code 
\gracklec\footnote{https://grackle.readthedocs.org/}$^,$\footnote{Changeset: 88143fb25480} 
\citep{Enzo_2014, Kim_2014}. \grackle self-consistently 
solves the 33 set reaction network including photo-ionisations. The network includes the most 
up-to-date rates as described in \citet{GloverJappsen_2007, GloverAbel_2008, GloverSavin_2009,
Coppola_2011, Coppola_2012,  Glover_2015a, Glover_2015b,  Latif_2015}. The reaction network
is described in full in R16.
The gas is allowed to cool radiatively during the simulation and this is also accounted for 
using the \grackle module. Here the rates have again been updated to account for recent updates 
in the literature \citep{Glover_2015a}. The cooling mechanisms included in the model are collisional
excitation cooling, collisional ionisation cooling, recombination cooling, bremsstrahlung and 
Compton cooling off the CMB. 

\subsection{Realisations}
In this study we compare two different realisations which we name Halo A and Halo B. Both halos were previously
determined in \cite{Regan_2015} and created with the \texttt{MUSIC} code. Using exactly the same methods as 
employed in R16 we place a radiating source (i.e. a ``galaxy'') close to a collapsing halo and investigate the 
effects of the realistic radiation field on the collapse of the halo and determine the viability of the 
direct collapse method. The idea that close-by neighbours are required for direct collapse has previously 
been studied analytically by \cite{Dijkstra_2008, Dijkstra_2014} and more recently using synchronised halo pairs 
by \cite{Visbal_2014b}. For each simulation we switch on the radiating source at a redshift of $z=40$ and place 
the source at a distance of 1 kpc, 2 kpc or 4 kpc physical from the target halo (i.e. point of 
maximum density at that redshift). We do not investigate sources for which the separation is less than 1 kpc 
as we found in R16 that this results in complete photo-evaporation of the halo. For each distance separation we 
also vary the spectrum of the radiating source. 
The spectrum is either a stellar only spectrum, a soft X-ray spectrum or a hard X-ray spectrum (see 
Figure \ref{StellarSED}). However, in all cases the spectrum is \textit{always} stellar for the first $\sim 20$
Myrs i.e. between a redshift of $z=40$ and $z=33$. At a redshift of $z = 33$ we either do nothing (stellar only 
case) or we update the spectrum to include soft X-rays (soft X-ray model) or we update the spectrum to include 
both soft and hard X-rays (hard X-ray model). The time between when the galaxy emits only stellar photons and 
when it begins to emit stellar plus X-ray photons is clearly uncertain. Our estimation of 20 Myrs takes into 
account the typical timescale of massive stellar evolution and the fact it takes time to build up a significant 
X-ray presence through binary evolution. In Table \ref{Table:radiation_particle} we have outlined 
each of the models used for our two realisations. The name of the simulation is made up as follows:
$<$InitialSeparation$>$\_$<$SpectraType$>$\_$<$Realisation$>$ where for SpectraType ``S'' stands for stellar only, 
``X'' stands for soft X-ray and ``HX'' stands for hard X-ray. 

%%%%%%%%%%%%%%%%%FIGURE 4%%%%%%%%%%%%%%%%%%%%%%%%%%%%%%%%%%%%%%%%%%%%%%%
\begin{figure}

 \center

  \includegraphics[width=9.5cm,height=8.5cm]{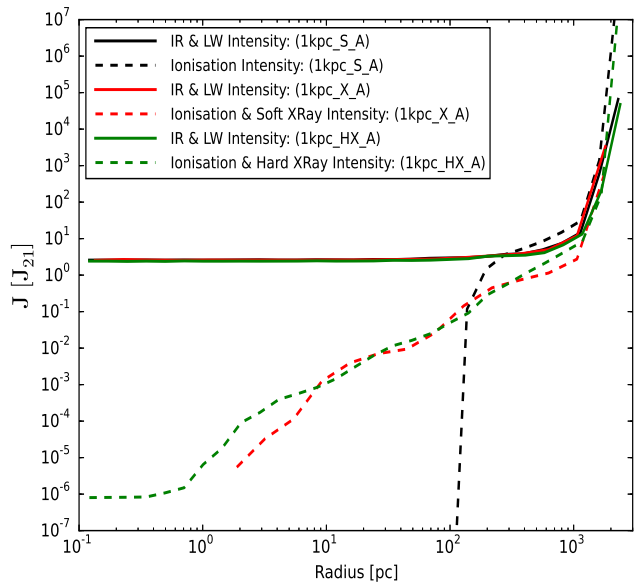}

  \caption{HaloA: The intensity ray profile for radiation emitted at a distance of 
          1 kpc (initially) from the collapsing halo. The intensity is broken into components below the ionisation
          threshold of hydrogen (IR \& LW) and that above the threshold. The black line refer to 
          the simulation using a Stellar spectrum only. The red line is from a simulation with a Stellar plus
          soft X-ray flux and the green line includes in addition also a hard X-ray flux. Radiation below 
          13.6 eV is able to penetrate deep into the halo with only minimal self-shielding. The ionising radiation 
          however suffers from varying degrees of absorption depending on the frequency of the radiation. 
          On the right hand axes we show the values of the ratio between the X-ray radiation and the 
          IR \& LW radiation for the hard X-ray spectrum model. The X-ray intensity is always sub-dominant
          to the IR \& LW radiation and drops sharply as the X-ray radiation is absorbed within approximately 
          100 pc of the centre. }

  \label{HaloAFluxProfile}

\end{figure}
%%%%%%%%%%%%%%%%%%%%%%%%%%%%%%%%%%%%%%%%%%%%%%%%%%%%%%%%%%%%%%%%%%%%%%%%%%%%

\section{Results}                   \label{Sec:Results}

\subsection{The impact of soft X-rays}

%%%%%%%%%%%%%%%%%FIGURE 5%%%%%%%%%%%%%%%%%%%%%%%%%%%%%%%%%%%%%%%%%%%%%%%
\begin{figure*}
  \centering 
  \begin{minipage}{175mm}      \begin{center}
      \centerline{
        \includegraphics[width=13cm]{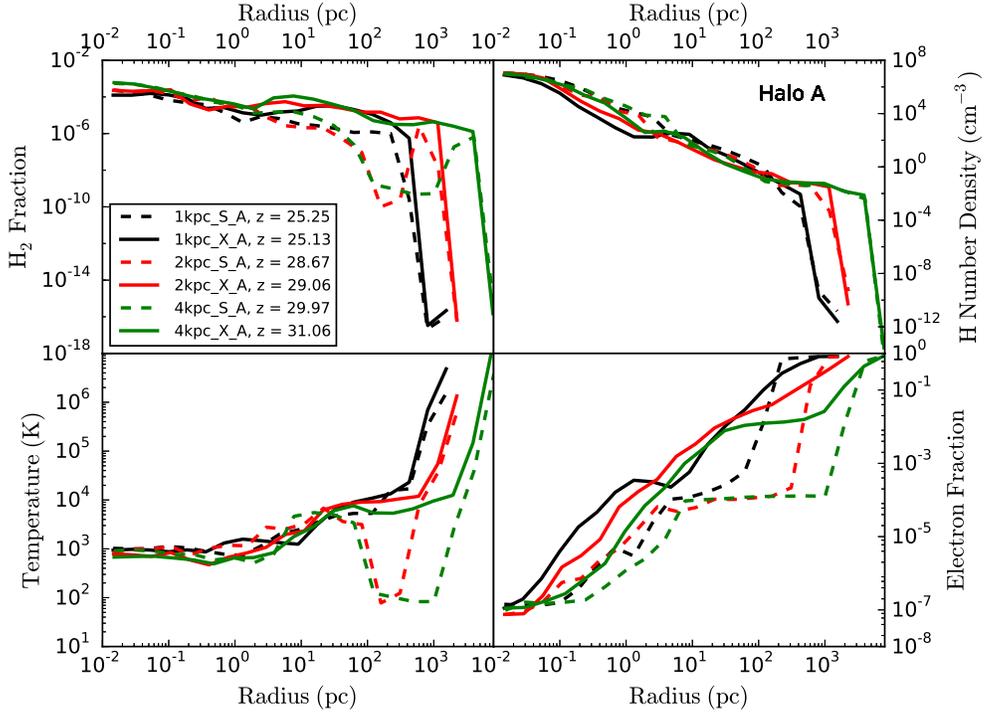}}        
        \caption[]
        {\label{HaloARayProfile} HaloA: Ray Profiles for Halo A at the final output time. The 
          panels starting from the top left and moving clockwise are: \molH Fraction, neutral hydrogen
          density, electron fraction and temperature. Included in each panel are simulations having a
          stellar only spectrum and those containing a stellar plus soft X-ray spectrum. The stellar 
          only simulations are indicated with an ``S'' suffix and those with the stellar plus soft X-ray 
          spectrum with an ``X'' suffix. Each profile is shown at the final output time.
        }
      \end{center} \end{minipage}
  \end{figure*}

%%%%%%%%%%%%%%%%%%%%%%%%%%%%%%%%%%%%%%%%%%%%%%%%%%%%%%%%%%%%%%%%%%%%%%%%%%%%

In order to properly assess the impact of the soft X-ray radiation component we break the analysis 
down into three constituent parts.  We begin by examining visually the impact of the X-rays.  We then 
analyse the impact of the X-rays by profiling the gas outwards from the point of maximum density back 
to the source, and finally we investigate the surrounding envelope of gas and look for effects at these larger 
scales. 

\subsubsection{Visual Inspection}
In Figures \ref{HaloBProjection_Stellar} and \ref{HaloBProjection_XRays} we show a projection of Halo B for 
first when the halo is exposed to a stellar spectrum only and then in the following plot when the Halo is 
exposed to a stellar plus soft X-ray spectrum. Visually Halo A and Halo B are very similar, we choose to show 
Halo B simply because there is more overall structure in the region surrounding this halo. The projections are 
made at the final output time in both cases. The first item to notice is that the 
gas is much hotter and also much more diffuse in Figure \ref{HaloBProjection_XRays} compared to Figure 
\ref{HaloBProjection_Stellar}. The soft X-ray component is able to heat more of the gas to higher temperatures
compared to the stellar only case. The gas in the model exposed to soft X-rays is also 
much more diffuse, looking at the gas number density projection shown in the bottom left panel 
there is an obvious lack of structure in the halo compared to the case where only a stellar spectrum is 
used. For the stellar model multiple high density structures exist with several density peaks clearly visible. \\ 
\indent The right hand panels of Figures  \ref{HaloBProjection_Stellar} and \ref{HaloBProjection_XRays} show the 
\molH fraction (top) and electron fraction (bottom), respectively. X-rays should produce more free electrons at 
larger scales because of their greater mean free paths while the ionising radiation will produce more free 
electrons local to the source. This is precisely what we see. \\

\subsubsection{Ray Profiles - Flux Statistics}
In Figure \ref{HaloAFluxProfile} we show the intensity in units of J$_{21}$\footnote{J$_{21}$ is defined as \J}. 
We define the intensity, $J$, exactly as we defined it in R16:
\begin{align}
J^\prime & = \sum_{E, i}  {k_i  E \over 4 \pi^2 \sigma_i(E)} \\
J  & = {J^\prime \over \nu_{H} J_{21}}
\end{align}
where $J^\prime$ is the sum of the intensities for each species, $i$, over all energy bins, 
$E$. Here $k_i$ is the number of photo-ionisations (or dissociations) per second for species $i$, 
$\sigma_i(E)$ is the cross section for species $i$ at energy $E$. Finally, $\nu_{\rm{H}}$ is the 
frequency at the hydrogen ionisation edge. The extra factor of $\pi$ in the denominator accounts for the 
solid angle.
The output is taken from Halo A when the initial separation is set to 1 kpc i.e. simulations 1kpc\_S\_A, 
1kpc\_X\_A  and 1kpc\_HX\_A. The profile is determined by averaging over 10 line of sight rays, each starting 
from the point source but each ray is given a small angular offset and so each ray travels along a slightly offset 
path to a circular region surrounding the point of maximum density. One of the 10 rays is exactly along a 
ray joining the source and point of maximum density, using a small number of rays means there is a weighting 
towards this line while still displaying an overall average. We break the radiation intensity into components 
below the ionisation threshold of hydrogen and those above the ionisation threshold. The solid lines show the 
radiation in the infrared (IR) and Lyman-Werner (LW) bands while the dashed lines show the radiation 
intensity for energies greater than 13.6 eV. The black line shows the intensity for the stellar only model,
the red line shows the intensity for the soft X-ray model while the green line shows the contribution 
from the hard X-ray model. The LW and IR intensities are identical in all cases as expected with a value of a 
few times J$_{21}$ in the core. This part of the spectrum is not affected by the inclusion of X-rays. However, 
the ionising components are quite different between the stellar and X-ray cases. The ionising radiation from the 
stellar source  is much less penetrating and is effectively blocked at a radius of $\sim 100$ pc. However, for 
the soft X-ray spectrum we are able to penetrate much more deeply into the
halo and in-fact can almost penetrate into the core of the halo - reaching down to a scale of $\sim 2$ pc. \\
\indent What is also clearly noticeable here is that the ionising intensity of the soft X-ray spectrum drops 
sharply as the rays penetrate into the halo and has fallen by approximately six orders of magnitude compared
to the IR \& LW intensities at small scales. In-fact over the range from a radius of 1000 pc down to $\sim 1$ pc 
the ionising intensity for the soft X-ray spectrum drops from an intensity of $\sim 1$ J$_{21}$ down to 
$\sim 10^{-6}$ J$_{21}$. The green line with triangles as markers shows the ratio of the X-ray intensity
($\mathrm{J_{X}}$) to the IR \& LW intensity ($\mathrm{J_{LW}}$) for the hard X-ray spectrum model. The values 
of the ratio are labeled on the right hand axes. The fall in the ratio of $\mathrm{J_{X}}$ to $\mathrm{J_{LW}}$ 
is clearly apparent as absorptions of the X-ray component take effect.
This is a direct consequence of both the 1/r$^2$ dependence of the radiation field and the 
impact of absorptions along the line of sight. The inclusion of hard X-rays does little to change the intensity 
values, the only significant impact of the hard X-rays is that they are able to penetrate to even smaller 
scales reaching well into the core of the proto-galaxy. \\
\indent When comparing the results found here with those elsewhere 
in the literature \citep[e.g.][]{Inayoshi_2011, Latif_2015, Inayoshi_2015b, Hummel_2015} it is important to 
bear this dependence in mind as other work has generally assumed a fixed relationship between the IR \& 
LW intensity and the ionising/X-ray intensity which is clearly not going to be the case for nearby sources. We will 
come back to this point in \S \ref{Sec:Discussion}.
%%%%%%%%%%%%%%%%%FIGURE 6%%%%%%%%%%%%%%%%%%%%%%%%%%%%%%%%%%%%%%%%%%%%%%%
\begin{figure*}
  \centering 
  \begin{minipage}{175mm}      \begin{center}
      \centerline{
        \includegraphics[width=13cm]{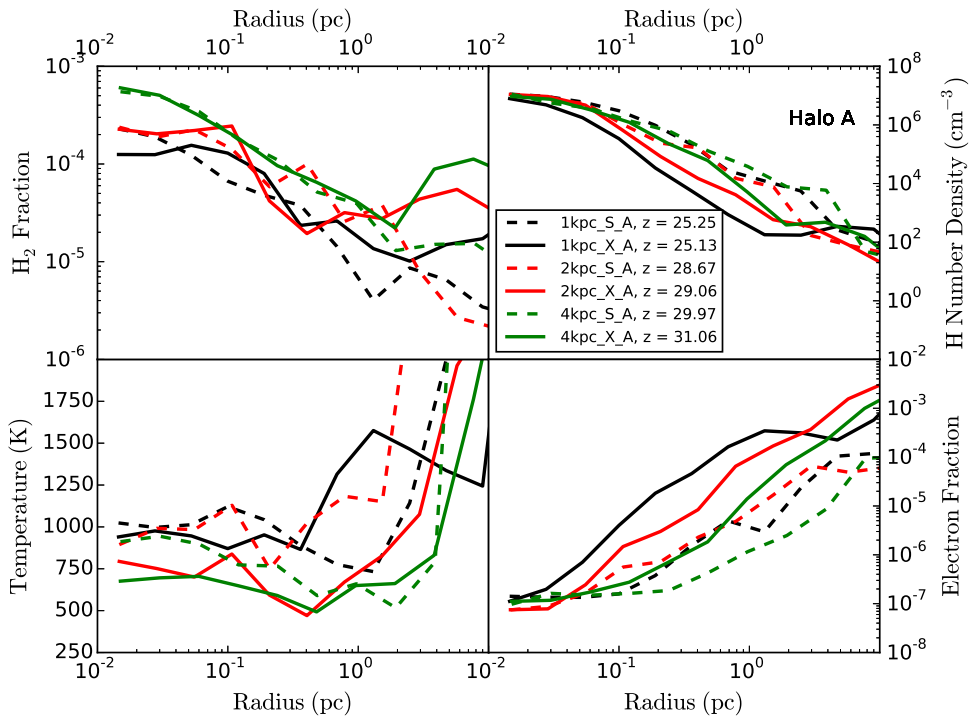}}
        
        \caption[]
        {\label{HaloARayProfile_Zoom} HaloA (Zoom): The same as Figure \ref{HaloARayProfile}
          except that the region of interest has been set to between 0.01 and 10 pc. The scale
          on the temperature plot has been changed to a linear scale on the y-axis so that the
          temperature in the centre of the halo is clearly seen and the impact of the different
          spectra more clearly identifiable. 
        }
      \end{center} \end{minipage}
  \end{figure*}

%%%%%%%%%%%%%%%%%%%%%%%%%%%%%%%%%%%%%%%%%%%%%%%%%%%%%%%%%%%%%%%%%%%%%%%%%%%%

%%%%%%%%%%%%%%%%%FIGURE 6%%%%%%%%%%%%%%%%%%%%%%%%%%%%%%%%%%%%%%%%%%%%%%%
\begin{figure*}
  \centering 
  \begin{minipage}{175mm}      \begin{center}
      \centerline{
        \includegraphics[width=13cm]{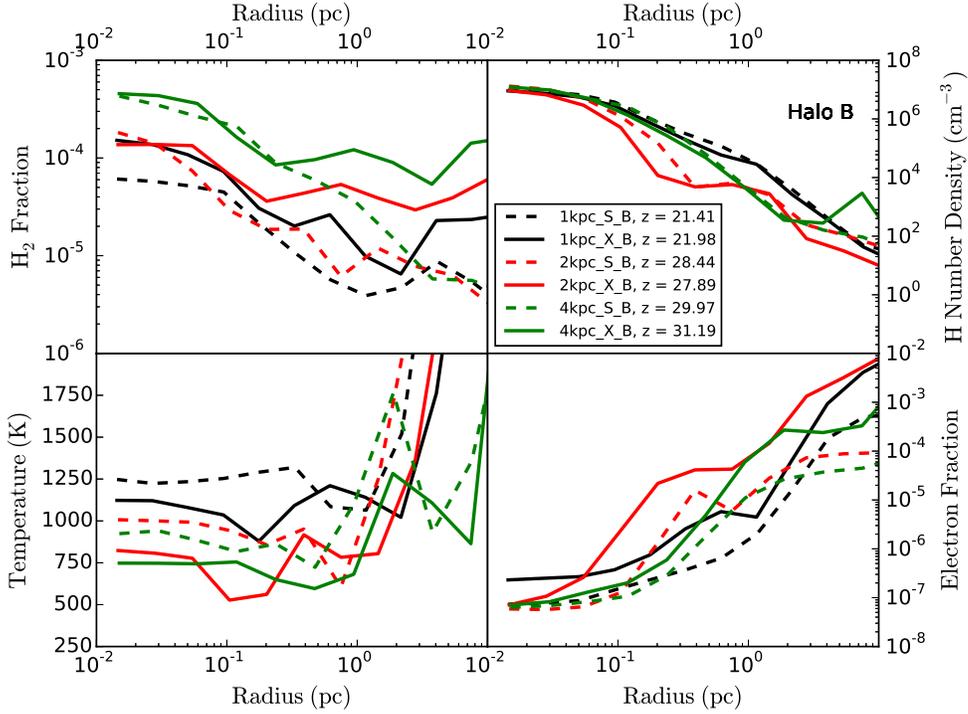}}
        
        \caption[]
        {\label{HaloBRayProfile} HaloB (Zoom): A zoomed in ray profile for Halo B with the horizontal radius 
          scale again set to between 0.01 and 10 pc. Similar to Figure \ref{HaloARayProfile_Zoom}.
        }
      \end{center} \end{minipage}
  \end{figure*}

%%%%%%%%%%%%%%%%%%%%%%%%%%%%%%%%%%%%%%%%%%%%%%%%%%%%%%%%%%%%%%%%%%%%%%%%%%%%

\subsubsection{Ray Profiles - Thermal Characteristics} \label{Sec:RayProfiles}
We now compare the profiles of the gas systematically across a broad range of realisations. 
In Figure \ref{HaloARayProfile} we have plotted ray profiles for Halo A for the case of the stellar spectrum and
the soft X-ray spectrum. In this case 1000 rays are used to construct the profiles. If we begin by examining the 
temperature plot (lower left panel) we can see that the solid curves depicting runs with soft X-rays 
all show a significantly higher temperature at scales greater 
than $\gtrsim 100 \ \rm pc$. The solid curves are those due to the soft X-ray spectrum and so the higher 
temperatures are due to the increased heating effects of the X-rays. At smaller scales the differences between 
the simulations are difficult to identify and we will inspect this region more closely in Figure 
\ref{HaloARayProfile_Zoom}. Looking next at the top left panel the \molH fraction is consistently higher for 
the simulations which include
a soft X-ray component. This can be understood in terms of the gas chemistry, the X-rays induce more ionisations
thereby increasing the free electron fraction (see the lower right panel for confirmation of this) which generates
more \molH via the two step Solomon process. The top right panel shows the neutral hydrogen density and agrees
well with what we saw in Figure  \ref{HaloAFluxProfile}. \\
\indent To get a better quantitative picture of what the impact of the soft X-rays is on the central object 
forming at the centre of the halo we now zoom into the central 10 pc region and examine the same quantities 
at smaller scales where the differences in the spectrum may impact on what type of object could finally form 
in such a region. In Figure \ref{HaloARayProfile_Zoom} we show the region within 10 pc for Halo A while in 
Figure \ref{HaloBRayProfile} we show the same region for Halo B. All of the ray profiles are created from the 
final output time. Rigid systematic differences are not obvious as both the distance is changed and the spectrum 
is changed from stellar to stellar plus X-rays. However, some trends are nonetheless still clear:
\begin{itemize}
\item For the stellar spectrum only, as the distances are decreased the temperature in the centre increases 
in both cases albeit more for Halo B ($\sim 30$\%) than Halo A ($\sim 10$\%). This is because the  \molH 
fraction is highest in the cases where the radiation source is furthest from the collapsing halo. 
This is an obvious consequence of the r$^{-2}$ dependence of the LW radiation field. Less \molH is destroyed 
by the sources which are further away.
\item When the X-rays are included, the temperature in the core in all cases decreases by approximately 10\%. 
The \molH fractions in the core are comparable for Halo A between the stellar and X-ray case while for Halo B
the \molH fractions are higher for the X-ray case. The higher \molH fractions does, at least for Halo B, induce 
some extra cooling in the core as a result.
\item We do not find that soft X-rays cause the halos to collapse earlier as a general rule. When comparing the 
impact of soft X-ray radiation to stellar radiation we find that in 2 out of 6 cases the halo collapses later. 
Naively one might expect the X-rays to generate more \molH at low and intermediate densities which overcomes any
heating effects to promote an earlier collapse time (compared to the stellar only case). However, we find this is
not always true and rather the complex interplay between X-ray heating, \molH formation, LW photo-dissociation 
and IR photo-detachment means that the collapse and also the collapse time is somewhat chaotic. However, 
as we will see explicitly later the X-rays do result in less massive cores.
\end{itemize}
Outside of 1 pc the \molH fraction for the cases where the X-rays are included can easily be an order of 
magnitude higher when compared to the stellar only case. However, as we profile into the core of 
the halo these differences become less pronounced 
and the \molH fractions tend to converge towards the stellar only result. However, the convergence is 
not perfect and differences can exist between the stellar result and the soft X-ray result. This
is clearest in the 1kpc\_X\_B case where the \molH differs by a factor of more than two in the centre 
between the two spectra - although this still only leads to a temperature difference of the order of 10\%. 
%%%%%%%%%%%%%%%%%FIGURE 7%%%%%%%%%%%%%%%%%%%%%%%%%%%%%%%%%%%%%%%%%%%%%%%
\begin{figure*}
  \centering 
  \begin{minipage}{175mm}      \begin{center}
      \centerline{
        \includegraphics[width=9cm]{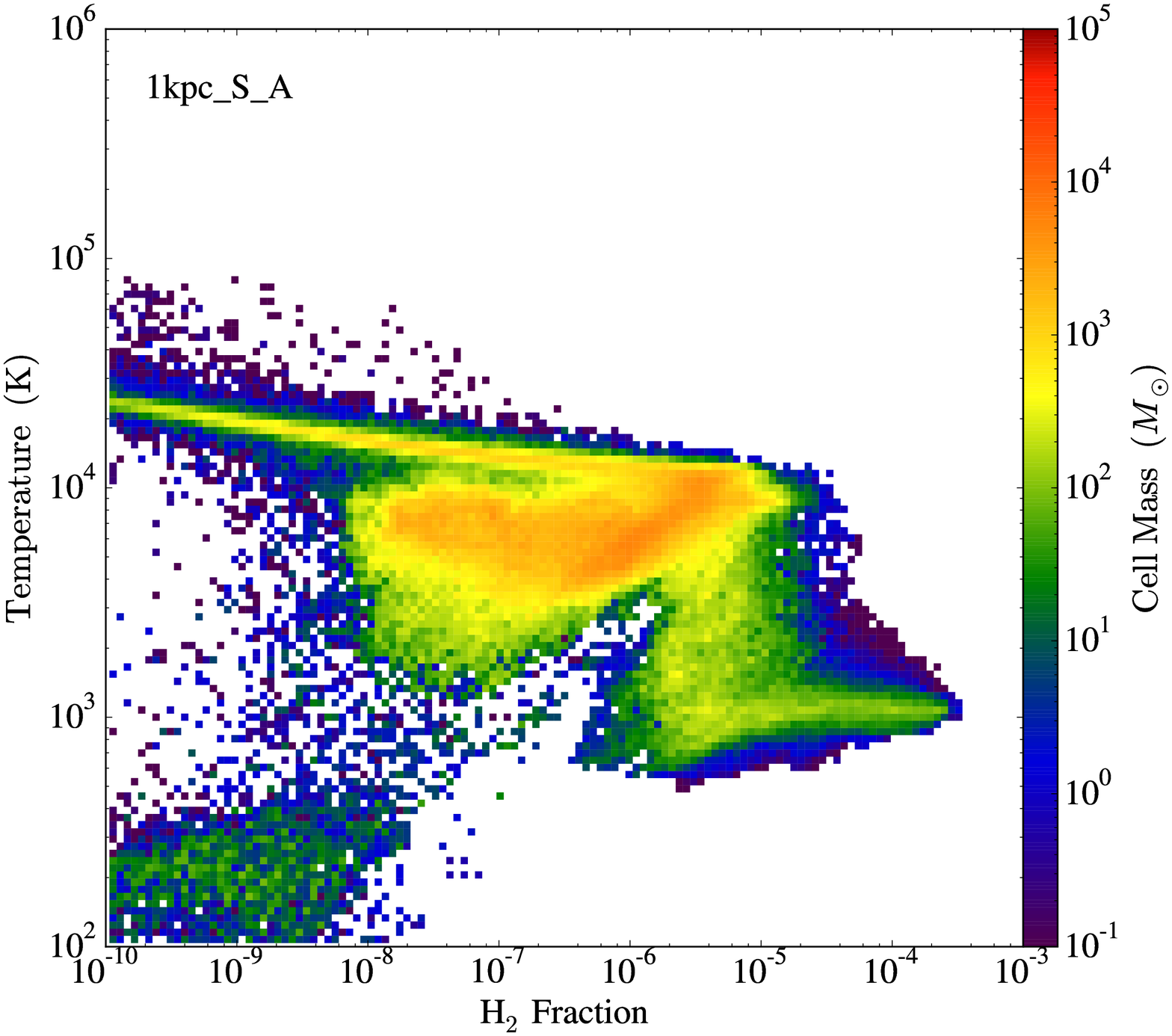}
        \includegraphics[width=9cm]{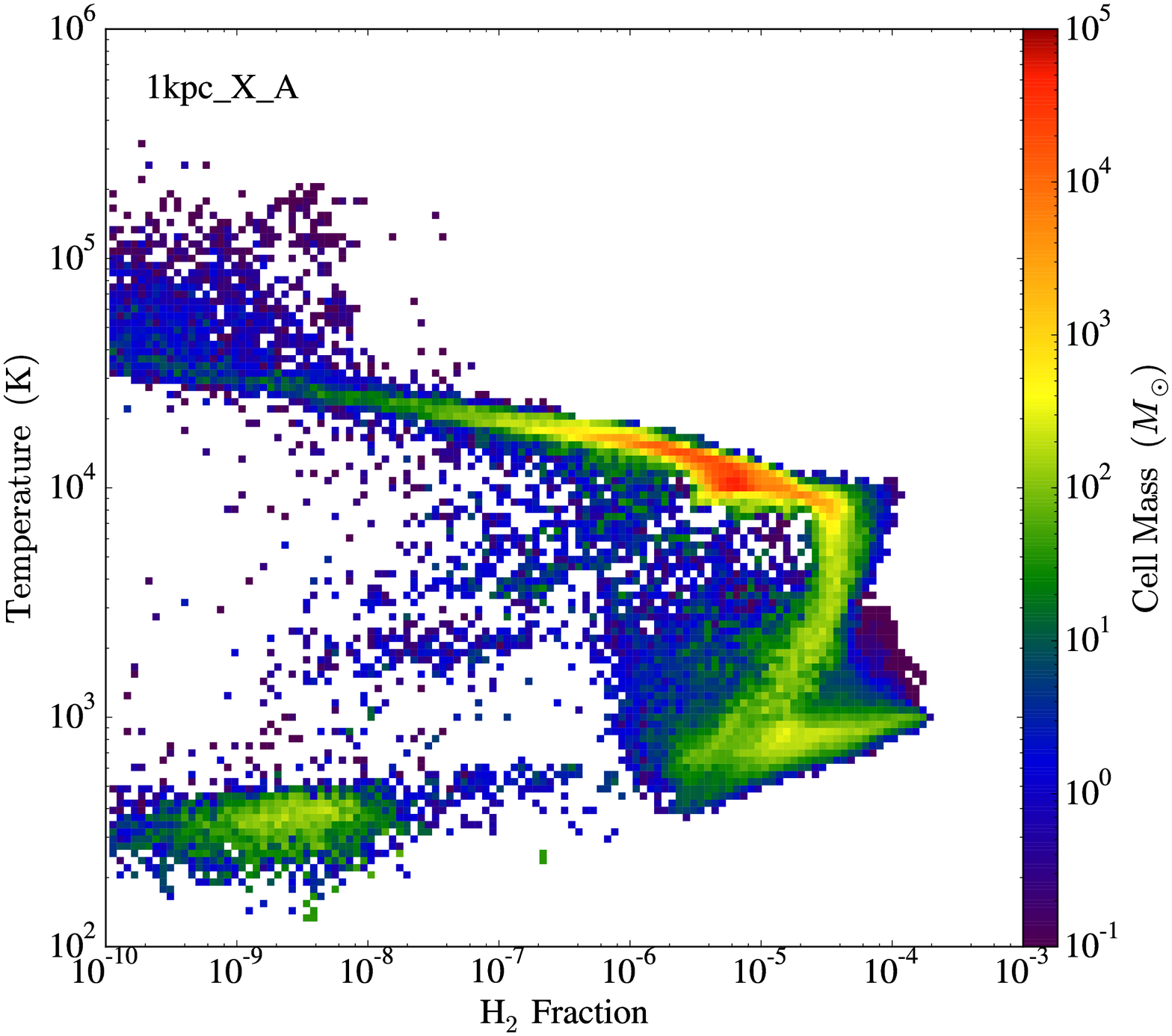}}
        \caption[]
        {\label{PhaseHaloA} Halo A: Phase diagram of \molH fraction versus temperature weighted by enclosed 
          cell mass. The left hand panel is for the stellar only model at an initial separation of 1 kpc, the right 
          hand panel for the stellar plus soft X-ray model at an initial separation of 1 kpc. The X-rays 
          produce a tighter relationship between \molH and temperature by heating the gas and not allowing 
          the gas to cool as efficiently forcing the gas to remain on the atomic cooling track until higher 
          \molH are reached. The gas masses in the bottom left corner of each plot is low density gas beyond 
          the edge of the HII regions which is cool and has a depleted \molH fraction.
        }
      \end{center} \end{minipage}
  \end{figure*}

%%%%%%%%%%%%%%%%%%%%%%%%%%%%%%%%%%%%%%%%%%%%%%%%%%%%%%%%%%%%%%%%%%%%%%%%%%%%
%%%%%%%%%%%%%%%%%FIGURE 8%%%%%%%%%%%%%%%%%%%%%%%%%%%%%%%%%%%%%%%%%%%%%%%
\begin{figure*}
  \centering 
  \begin{minipage}{175mm}      \begin{center}
      \centerline{
        \includegraphics[width=9cm]{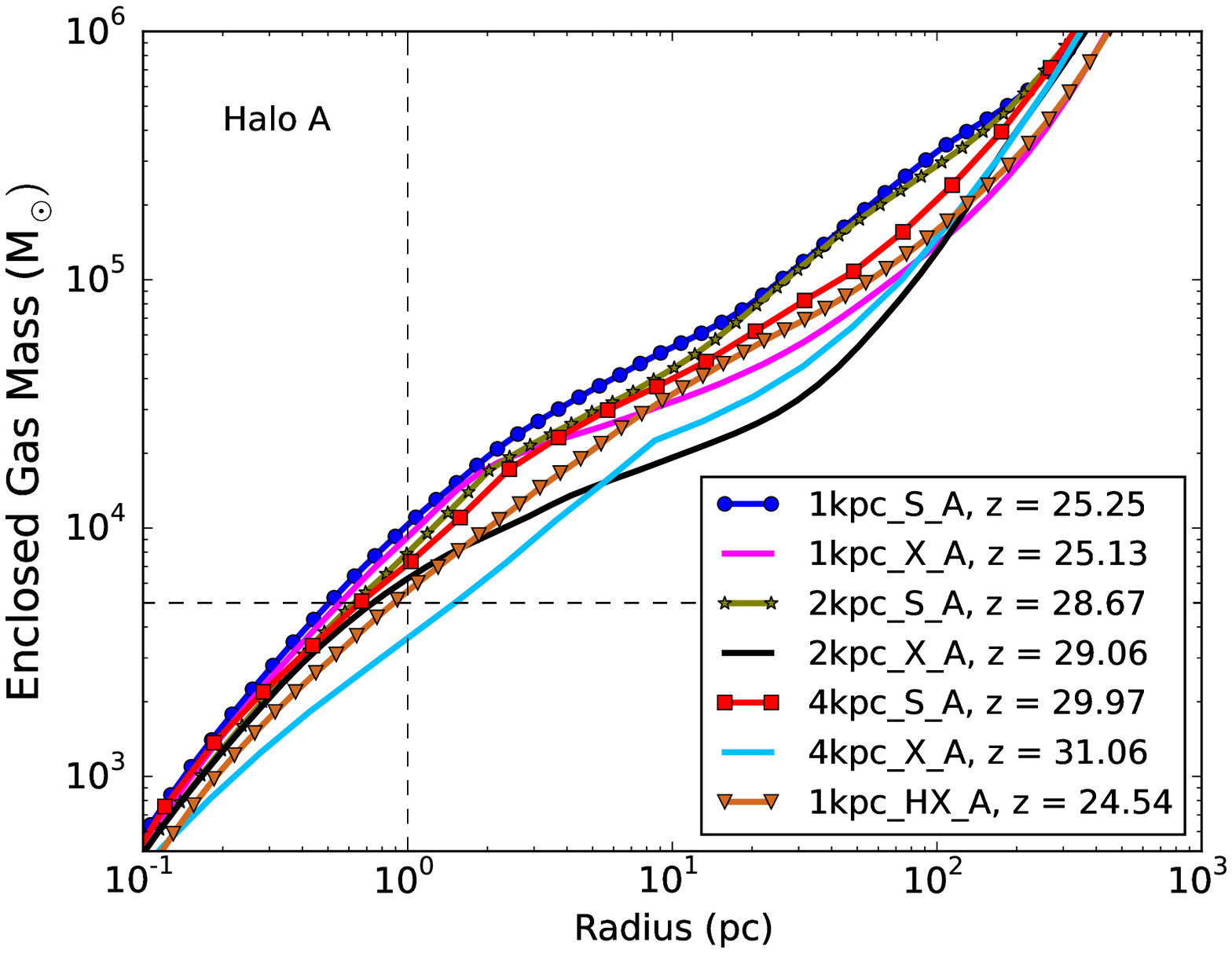}
         \includegraphics[width=9cm]{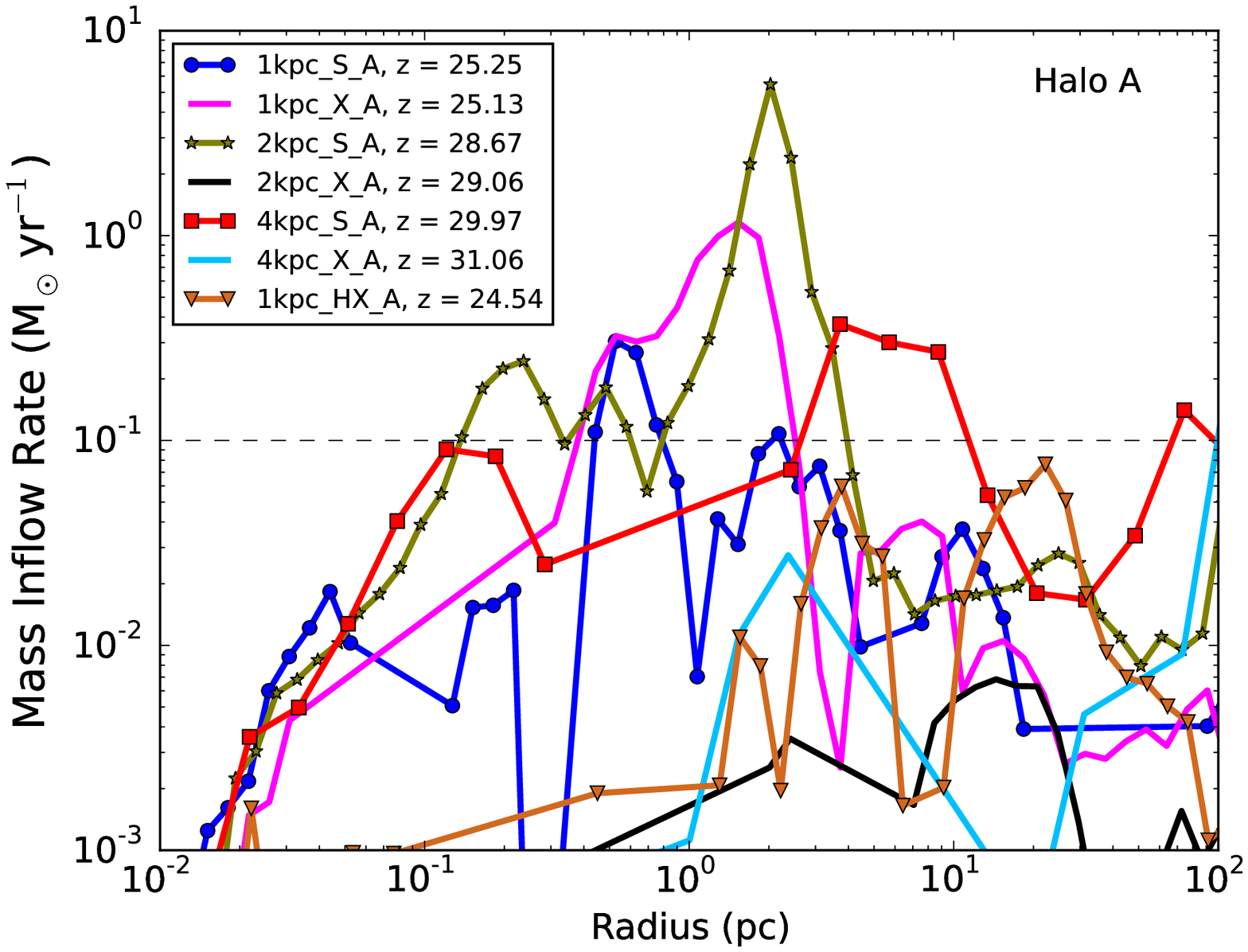}}
        \caption[]
        {\label{HaloARadialProfiles}
        Radial Profiles for HaloA.  The left hand panel shows the enclosed mass profile from 0.1 pc up to 
        1000 pc. The deleterious effects of the X-rays are most noticeable in this case for the 
        models in which the initial separation is greater than 1 kpc, in the 2kpc\_X\_A and 4kpc\_X\_A the halo 
        collapses earlier and the enclosed mass is reduced significantly. The right hand panel shows 
        the accretion rates from 0.01 pc to 100 pc out from the maximum density. The dashed line at a mass 
        inflow rate of 0.1 \msolar yr$^{-1}$, is shown as approximately the mass inflow rate required to produce 
        a super-massive star.

        }
      \end{center} \end{minipage}
  \end{figure*}

%%%%%%%%%%%%%%%%%%%%%%%%%%%%%%%%%%%%%%%%%%%%%%%%%%%%%%%%%%%%%%%%%%%%%%%%%%%%
%%%%%%%%%%%%%%%%%FIGURE 10%%%%%%%%%%%%%%%%%%%%%%%%%%%%%%%%%%%%%%%%%%%%%%%
\begin{figure*}
  \centering 
  \begin{minipage}{175mm}      \begin{center}
      \centerline{
        \includegraphics[width=13cm]{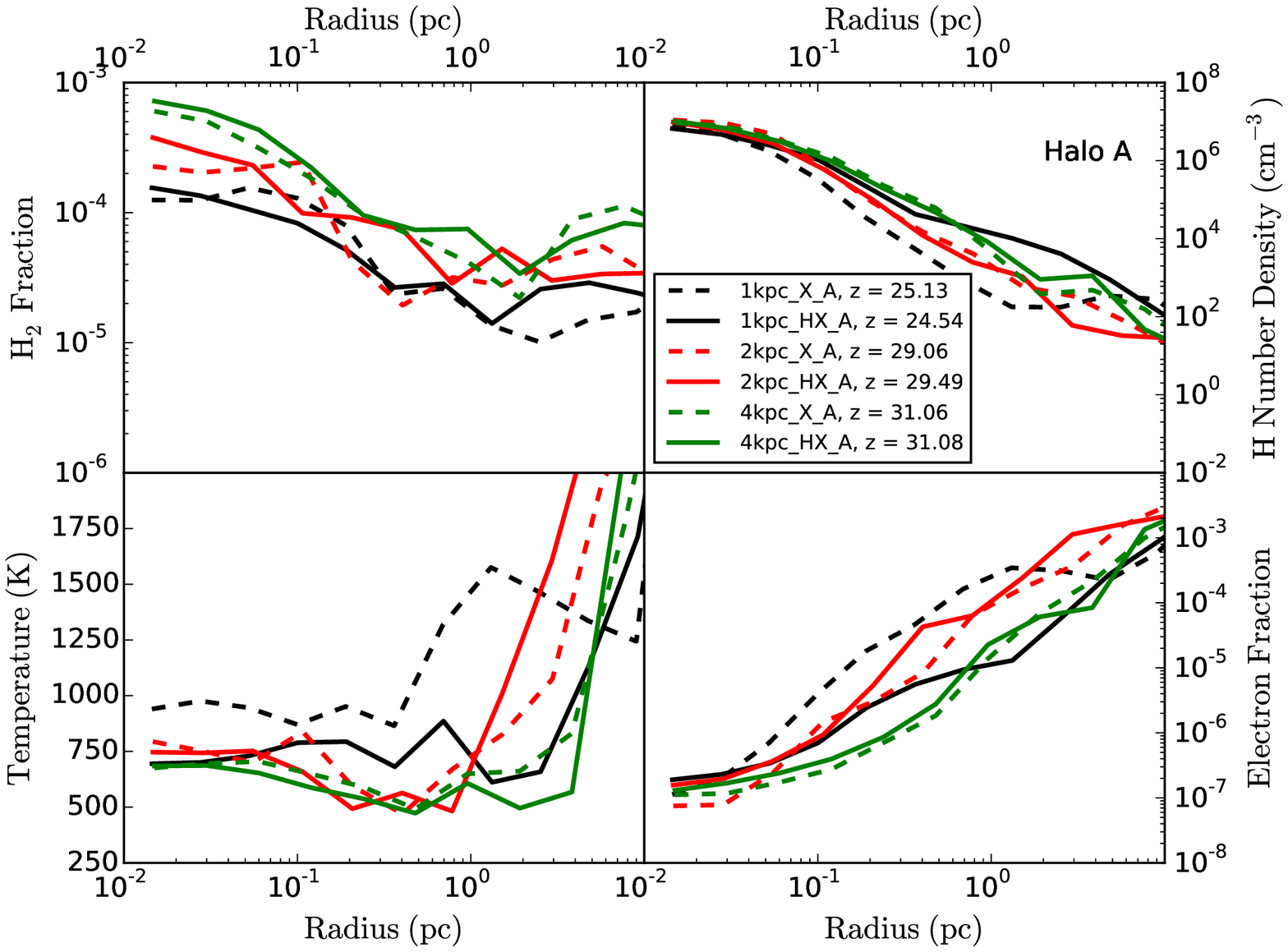}}
        
        \caption[]
        {\label{HaloARayProfile_HardXRays} HaloA (HardXRays): This figure shows ray profiles for the 
          cases where soft X-ray and hard X-ray models are used. The hard X-ray models differ only in that 
          photons with energies $\gtrsim 1$ keV are included in the model. The inclusion of hard X-rays has only 
          a small effect on the gas quantities for the cases where the separation is greater than 1 kpc. For the 
          1 kpc realisation the temperature is approximately 300 K lower in the hard X-ray case compared to the 
          soft X-ray case but it is consistent with the other profiles. 
        }
      \end{center} \end{minipage}
  \end{figure*}

%%%%%%%%%%%%%%%%%%%%%%%%%%%%%%%%%%%%%%%%%%%%%%%%%%%%%%%%%%%%%%%%%%%%%%%%%%%%
\subsection{The Surrounding Envelope and Accretion Rates}
In Figure \ref{PhaseHaloA} we examine the distribution of \molH as a function of temperature weighted by 
cell mass. We only show the results from Halo A as the results from Halo B are qualitatively very similar. 
In the left hand panel 
we show the output from Halo A at the final output time when only a stellar spectrum is used, in the right 
hand panel we show
the final output time for the case of a stellar plus soft X-ray spectrum. Visually the difference are quite 
striking, the stellar only model has a much broader distribution of gas in terms of temperature and to a 
lesser extent in the \molH fraction. The stellar model has a large mass of gas between $T\sim 10^3\ \rm K$ and 
$T\sim 10^4 \ \rm K$ with a \molH fraction between $10^{-8}$ and $10^{-5}$. 
The model including X-rays however has much narrower temperature distribution with most of the gas sitting 
at $T\sim 10^4\ \rm K$ even though the \molH fraction is actually higher at values between $10^{-7}$ and $10^{-4}$. 
However, the heating effects of the X-rays at this close separation means that the bulk of the gas is heated 
to $10^4$ K with the increased \molH fraction and the increased associated cooling being unable to counteract 
the heating effect. \\
\indent This increased temperature of the gas when exposed to X-rays, most especially the gas at scales 
greater than 10 pc, means that the enclosed mass fraction is always higher at a given scale for gas exposed to 
a stellar only spectrum compared to an X-ray spectrum. In the left hand panel of Figure \ref{HaloARadialProfiles}
we show the enclosed mass as a function of radius for Halo A. The enclosed mass is greatest when the source is 
closest to the collapsing halo and when it is exposed to stellar photons only. X-rays show a systematic reduction 
in the enclosed mass when compared to the stellar spectrum which becomes more pronounced as the distance to
the source increases. This is because the LW radiation disrupts \molH cooling effectively when the flux is 
strongest (closest) and the ionising radiation is not as efficient at heating the gas compared to X-rays 
at these scales. Hence, the \molH fraction is lowest when the source is closest and for the stellar spectrum 
resulting in a larger enclosed mass collapsing. The same mechanism also has an effect on the mass in-flow rates, 
albeit weaker, as shown in the right hand panel of Figure \ref{HaloARadialProfiles}. For the larger separations 
with X-rays we see that the mass inflow rates are quite low. This is because the X-rays heat the gas reducing 
its ability to cool and thus leads to lower infall rates. However, at a 1 kpc separation the accretion rate, 
even in the X-ray case, is very high. In fact for the X-ray case at a separation of 1 kpc the peak mass inflow 
rate exceeds the stellar case. What is clear is that the increased temperature of the gas, compared to the 
stellar case, results in a reduced mass inflow rate and that X-rays at separations of 2 kpc or more give the 
lowest mass inflow rates. However, when the separation drops to 1 kpc the negative \molH inducing impact of the 
X-rays disappears and the impact of the X-rays becomes neutral and may even switch sign to being marginally 
positive. \\
\indent The formation of SMS is postulated when the accretion rates onto a central object can exceed 
$\sim 0.1$ \msolar yr$^{-1}$ \citep{Begelman_2006, Johnson_2012, Hosokawa_2013, Schleicher_2013}. Our mass 
inflow rates peak at values much larger than $0.1$ \msolar yr$^{-1}$  for the nearby radiation sources. 
Assuming a lifetime of $\sim 1$ Myr for such a massive star and an initial mass of $\rm{M_{init}} \sim 10^4$ 
\msolar the star could grow to a mass exceeding a few times $10^5$ \msolar by the end of its short lifetime. 
More in-depth simulations, which are beyond the scope of this study, of the continued evolution of this 
particular collapse would be required to support this hypothesis. Such a simulation would need to include 
detailed stellar evolution modelling of SMS formation \citep[e.g.][]{Hosokawa_2011, Hosokawa_2012, Hosokawa_2013, 
Inayoshi_2014}.

\subsection{Does a Hard X-ray Spectrum Make Any Difference?} \label{Sec:HardXRays}
\begin{figure*}
  \centering 
  \begin{minipage}{175mm}      \begin{center}
      \centerline{
        \includegraphics[width=9cm]{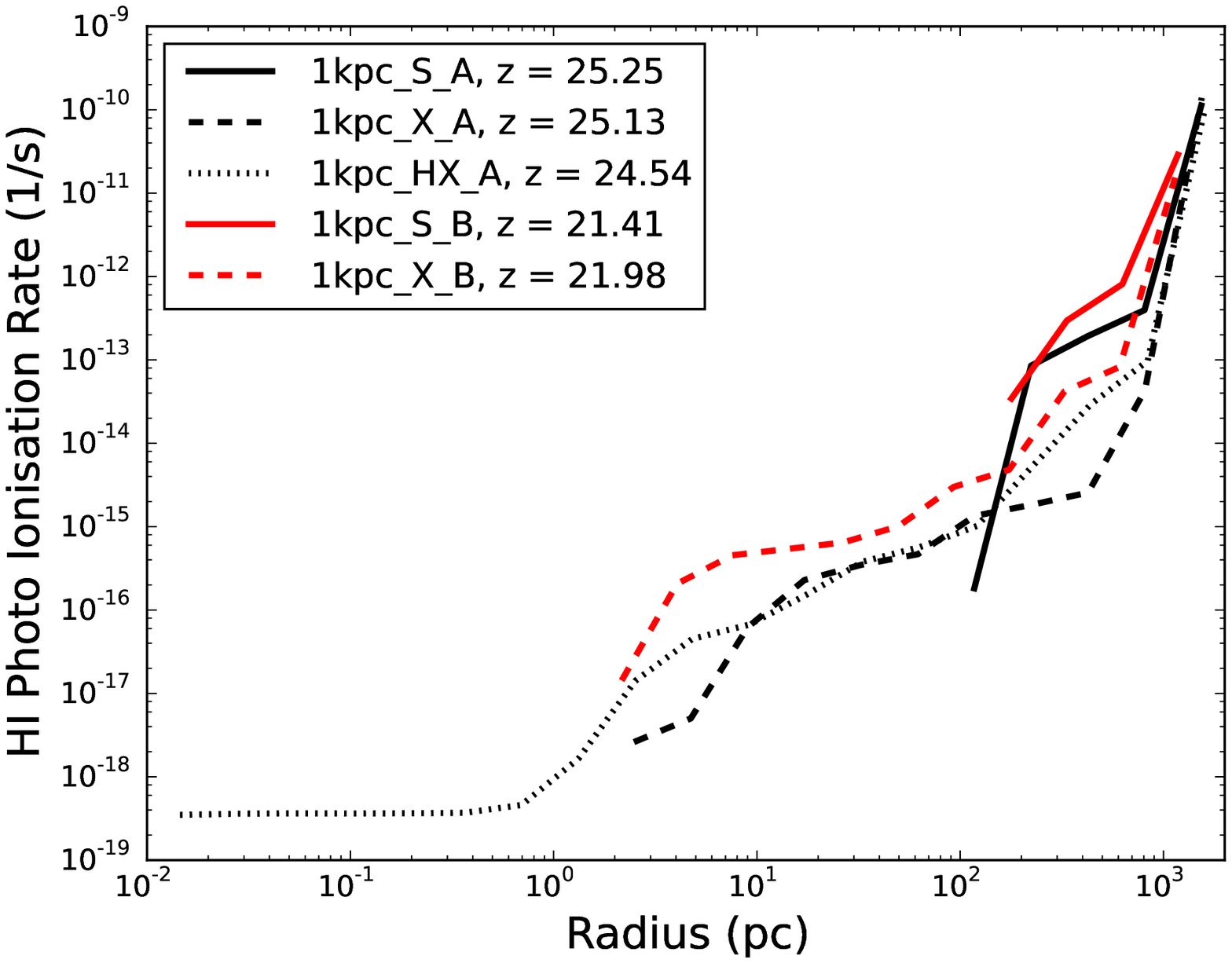}
        \includegraphics[width=9cm]{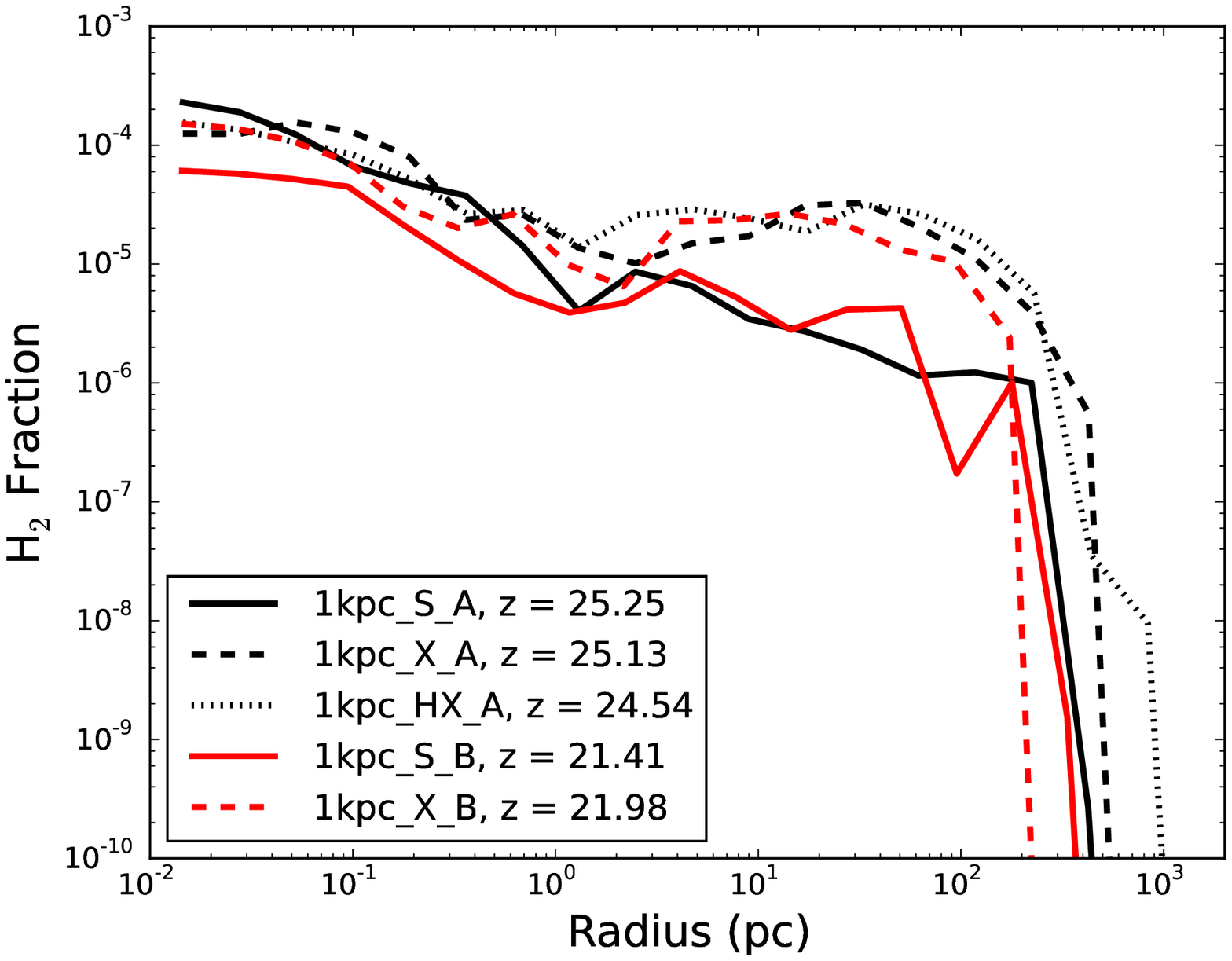}}
        \caption[]
        {\label{RayProfiles_1kpc} 1 kpc models: The two panels show ray profiles for all of the models
          where the initial separations is 1 kpc. Black for Halo A and red for Halo B. Solid lines for 
          the stellar spectra, dashed for the soft X-rays and dotted for the hard X-ray models. In the left
          hand panel we see that the models including X-rays show significantly more ionising ability. In the 
          right hand panel the strong increase in \molH fraction between 3 \& 300 pc is clear for the X-ray case.
          However, this increased \molH fraction does not lead to a temperature reduction as the heating effects 
          of the X-rays dominate at these densities. 
          
        }
      \end{center} \end{minipage}
  \end{figure*}

In Figure \ref{HaloARayProfile_HardXRays} we show the impact of hard X-ray photons on the gas state when 
compared to the soft X-ray models. The hard X-ray models are described in Table \ref{Table:radiation_particle}. 
The hard X-ray models increase the number of energy bins required from 8 to 11 and the subsequent runtime 
increases significantly (the 1kpc\_HX\_A run took more than 60 days wall-clock time to complete ($\sim 370,000$
CPU hours) compared to an average runtime of 10 days ($\sim 62,000$ CPU hours)). As a result the 
hard X-ray model was only run for Halo A. The mean free path of the hard X-rays is longer than for the 
soft X-rays as their interaction cross section is smaller. This feature is also confirmed in 
Figure \ref{HaloAFluxProfile} where we see that the intensity due to hard X-rays is almost identical to soft 
X-rays but with a deeper penetration (this was for an initial separation of 1 kpc in each case). \\
\indent In the bottom left panel of Figure \ref{HaloARayProfile_HardXRays} we see that hard 
X-rays (solid lines) have little impact on the temperature of the gas compared to the soft X-ray case for the 
2 kpc and 4 kpc cases. For the case of the 1 kpc separation the temperature of the gas in the core of the halo
is approximately 300 K lower compared to the soft X-ray case. We have over-plotted the enclosed 
gas mass and mass inflow rates for the 1kpc\_HX\_A runs in Figure \ref{HaloARadialProfiles}. It is clear from this 
figure that the enclosed mass values for the 1kpc\_HX\_A run is much lower than both the 1kpc\_S\_A and 
1kpc\_X\_A runs at distances up to $\sim 100$ pc from the centre. This trend is confirmed by the mass inflow 
values in the right hand panel. The reason for the reduced enclosed mass values is due to the variation in the 
penetrating ability of the photons as a function of their energies. More energetic photons are able to ionise 
the hydrogen to greater depths, suppression gas accretion and reducing the enclosed mass. \\
\indent As a result we see 
higher enclosed masses for the stellar only case compared to the soft X-ray case, for which the masses are again 
higher when compared to the hard X-ray case
at a radius of $\lesssim 10$ pc. The effect is somewhat cumulative, while soft X-rays do certainly induce a small 
negative effect here the hard X-rays enhance it to significant levels. \\
\indent We have explicitly compared this effect in Figure \ref{RayProfiles_1kpc} 
where we have taken the 1 kpc models and compared them as their spectrum is varied. We saw in Figure 
\ref{HaloARadialProfiles} that the enclosed 
mass values are connected to the penetrating ability of the ionising photons. In the left hand panel of 
Figure \ref{RayProfiles_1kpc} we see that the stellar spectrum photons 
get halted at a radius of $\gtrsim 100$ pc, the soft X-ray photons at closer to 1 pc and the hard X-ray photons 
make it all the way into the core. It is the extra ionisation caused by the hard X-rays which further suppresses 
the mass inflow rate compared to the soft X-ray and stellar case and hence the enclosed mass. \\
\indent \cite{Latif_2015} investigated the impact of hard X-rays photons (uniform background X-ray intensities of 
between J$_{X}$ = 0.01 and J$_{X}$ = 1.0)\footnote{ $\rm{J_{X} = J_{X, 21} \Big( {\nu \over \nu_0} \Big)^{-1.5}}$ 
and  $J_{X, 21 }$ is the Cosmic X-ray background flux in units of J$_{21}$ } and 
found that the hard X-rays increase the value of J$_{crit}$ by a factor 
of between 2 and 4. Their values of the X-ray intensities are significantly beyond what we simulate here, and 
more appropriate for the X-ray spectrum expected for nearby accreting super-massive black hole. \\
\indent In summary we find that hard X-rays from realistic sources have an additional negative effect 
compared to soft X-rays. Their ability to penetrate deep into a halo and ionise hydrogen leads to less centrally 
concentrated gas clouds, leading to lower core masses. 

%%%%%%%%%%%%%%%%%FIGURE 11%%%%%%%%%%%%%%%%%%%%%%%%%%%%%%%%%%%%%%%%%%%%%%%

%%%%%%%%%%%%%%%%%%%%%%%%%%%%%%%%%%%%%%%%%%%%%%%%%%%%%%%%%%%%%%%%%%%%%%%%%%%%

\section{Discussion}  \label{Sec:Discussion}
Disrupting or preventing completely the formation of \molH is seen as a necessary criteria for the direct 
collapse model of SMBH formation. As a result nearby, strongly luminous, galaxies which produce copious amounts 
of Lyman-Werner radiation are seen as a vital component. It is however, also clear that these galaxies will 
form at least some HMXBs which will lead to a significant X-ray component on top of the stellar component. In 
this work we have investigated thoroughly the added impact of both soft and hard X-rays compared to a stellar 
only spectrum. \\
\indent There has been some debate in the literature as to the feedback effects of X-rays on SMBH formation. 
\cite{Hummel_2015} investigated the effect of Population III star formation under X-ray feedback. They found 
that the gas becomes optically thick to X-rays at densities above approximately n$_H \sim 10^4$ cm$^{-3}$ and 
that as a result Pop III star formation is relatively insensitive to the presence of a cosmic X-ray background. 
\cite{Inayoshi_2015} came to a slightly different conclusion in the context of direct collapse black holes. 
They found that the impact of soft X-rays is to increase the value of J$_{crit}$ thus making DCBH formation less 
likely. In their study they set the intensity of X-rays to approximately $10^{-5}$ times that of the LW intensity 
(see their equation 14). They find that the critical LW intensity required for direct collapse is increased by at 
least an order of magnitude when X-ray intensities of $\mathrm{J_{X}} \gtrsim 0.01$ are included. However, 
their results are not for a single source and instead they consider a much larger far ultra-violet flux 
(which could be due to multiple nearby halos) and scale the X-ray flux proportionately. As such they investigate 
a somewhat different scenario to that of a single dominant source. By comparison we evolve the radiation 
field self-consistently in 3-D. In Figure \ref{HaloAFluxProfile} 
we see a very strong decrease in the X-ray intensity compared to the LW intensity as we move towards the 
centre of the collapse. It is this variation in the X-ray intensity with distance that will ultimately 
determine the feedback effects from the X-rays as we discuss below. \\
\indent Our detailed modelling shows that (similar to \citealt{Hummel_2015}) the inner regions of the halo are 
agnostic to the X-rays and hence the thermal characteristics of the gas are relatively insensitive to the 
X-ray component. We see only small changes in the thermal characteristics of the core of the halo with the 
inclusion of X-rays. The impact is especially small when the initial separation is small and only grows 
slightly as the X-ray source is moved further away. \\
\indent However, X-rays do have a significant effect on the gas surrounding the core i.e. gas between 1 pc and 
a few times $10^2$ pc from the central maximum. As the X-ray source is moved 
further from the halo we see the gas in the envelope surrounding the core is negatively affected.  The negative 
feedback effects of the X-rays are seen clearly in terms of the enclosed 
mass of the halo and more weakly, in the mass inflow rates. This distance dependence can be understood 
in terms of the effect of the X-rays on the low and medium density gas in particular (i.e. gas at a density of
n$_{H} \lesssim 10^2$ cm$^{-2}$). The X-rays, compared to the stellar only case, result in more diffuse gas which is 
much hotter than the gas in the stellar only case (see Figure \ref{HaloARayProfile}). For the cases where the 
separation is 2 kpc and 4 kpc respectively the gas is approximately two orders of magnitude hotter in the X-ray 
case leading to significantly reduced accretion rates and hence smaller core masses.\\
\indent In the range $r = 3-300$ pc (see for example Figure \ref{RayProfiles_1kpc} right hand panel), 
the \molH fraction increases by an order of magnitude when X-rays impact the system, which partially ionize the 
outer parts of the halo. We can estimate the equilibrium \molH fraction by setting the \molH formation time 
$t_{\rm form} \approx f({\rm H}_2) / k_{{\rm H}^-} n_b f(e^-)$ to its dissociation time 
$t_{\rm diss} = k_{\rm diss}^{-1} = 23/J_{21} \; \textrm{kyr}$ \citep{Yoshida_2003a, Wise_2007b}, arriving at
$f_{\rm eq}({\rm H}_2) \simeq (23 \; {\rm kyr} / J_{21}) k_{{\rm H}^-} \, n_b\, f(e^-)$.  Here $f(i)$ is the 
fractional abundance of species $i$, $n_b = 10 \ \rm{cm^{-3}}$ is the baryon number density and $k_{{\rm H}^-}$ is the 
H$^-$ formation rate coefficient by electron 
photoattachment and is around $10^{-15}$ cm$^3$ s$^{-1}$ at $T \approx 1000$ K \citep{Wishart_1979, GloverAbel_2008}.
Taking the conditions at $r \simeq 10$ pc in 1kpc\_X\_A, the equilibrium abundance $f_{\rm eq}({\rm H}_2) \simeq 2 
\times 10^{-5}$, in line with (or perhaps slightly above) the simulation data.  Because $f_{\rm eq}({\rm H}_2)$ 
scales with electron fraction, both the electron and \molH fraction drop by a factor of 10 in the stellar-only 
run. At this scale the electrons are in ionisation equilibrium. Comparing the recombination rate to the 
ionisation rate at the hydrogen edge leads us to an equilibrium value of $\rm{f(e^-)} \sim 4 \times 10^{-3}$ in 
the case of X-rays and $\rm{f(e^-)} \sim 7 \times 10^{-4}$ for the stellar case. The free electron fraction 
in the stellar case reaches a plateau (see Figure \ref{HaloARayProfile}) of $\rm{f(e^-)} \sim 1 \times 10^{-4}$  
between scales of $r \simeq 10$ pc and $r \simeq 1000$ pc which is its collisional equilibrium value as opposed 
to its photo-ionisation value seen in the case of X-rays. At any rate the heating effect of the X-rays 
dominates over any induced cooling effects from the enhanced \molH fraction. 
We see no material effect from the slightly elevated \molH fraction due to X-rays (compared to the stellar
only case), rather the heating effect dominates and suppresses the accretion rates. \\ 
\indent Inside of the cores, where densities are similar in both the stellar and X-ray cases the thermal 
characteristics are similar, the cores are simply less massive. 
For the case where the initial separation is 1 kpc the temperature profiles between 
the stellar and soft X-ray case are virtually identical leading to mass inflow rates which are very similar. 
In this case there is little negative impact due to the soft X-rays and in fact the mass inflow rates are 
slightly higher for the X-ray case. For a hard X-ray spectrum the photons can penetrate into the very core 
(see Figure \ref{RayProfiles_1kpc}). As a result hard X-rays induce a negative feedback effect at all 
separations, which is likely to be detrimental to (massive) star formation in halos exposed to such a spectrum.\\
\indent A significant caveat to our study is that we examine the case of a single radiation source. 
We do not attempt to model classical DCBH formation in this study, instead we focus solely 
on studying the effect of nearby (X-ray) radiation sources which are seen as a cornerstone of creating pristine
atomic cooling halos and by extension are a cornerstone of the DCBH formation mechanism.
While this allows us to disentangle the effects of a realistic radiation source from other nearby radiation 
sources it is unlikely to be the cosmologically realistic case.  As we clearly 
showed in \S \ref{Sec:RayProfiles} a nearby radiation source with characteristics similar to a first galaxy
is unable to fully dissociate \molH in a collapsing halo (the effect of this non-negligible \molH abundance 
on the gas thermo-dynamics is unclear - gas fragmentation may be one outcome - however, an investigation of the 
further evolution of the gas collapse is beyond the scope of this work). What will more likely be required 
is the scenario where a nearby source is augmented by additional sources clustered around rare density peaks. 
These additional sources will sum to produce a background radiation field which will for a given time be 
dominated by one (as simulated here), or at most a handful of nearby sources. Our work should therefore be seen 
as an initial test of the closely separated pairs mechanism  \citep{Visbal_2014b}. Our simulations 
show that a single nearby source will likely \textit{not} provide a sufficient condition for the formation of 
DCBH seed.\\
\indent A recent study by \cite{Chon_2016} uses the star particle technique together with a spatially and 
temporally varying LW radiation field including self-shielding to examine the conditions for direct collapse. 
They use a large volume (20 \mpch comoving) and include the effect of multiple sources finding
 multiple DC candidates. They conclude that while a nearby 
neighbour is required to provide a sufficiently intense LW radiation field the neighbour can also hamper the 
formation of a SMS through adverse dynamical interactions. In our study these dynamical effects are absent due to
our chosen setup. Furthermore, \cite{Chon_2016} find that the value of the LW intensity may not 
be as high as described more generally in the literature and may in fact be much lower than the often quoted value 
of $\rm{J_{crit} \sim 1000\ J_{21}}$ due to the presence of the near neighbour and the variation in the flux 
(and increase in the flux as the halos merge). We have specifically not simulated a nearby host with the 
intention of trying to uncover a single value for ``$\rm{J_{crit}}$'' but rather we focus on examining the case 
of a single galaxy with star formation rates and masses deemed likely at this redshift. \\

\section{Conclusions}  \label{Sec:Conclusions}
We have studied here the effects of X-ray feedback on forming direct collapse black hole seeds. 
Our conclusions are:
\begin{itemize}
\item \textbf{The incorporation of X-rays has a negligible effect on the thermal profile of the core of the halo.} 
The core of the halo feels only a very minimal effect from the X-rays due to self-shielding. At scales below 
approximately 1 pc the thermal profiles of all of our simulations look quite similar. The haloes irradiated by
X-rays do show small increases in the \molH fraction within the core and this does lead to a small reduction in 
the core temperature at the level of $\lesssim 10\%$ but the overall effect is small.
\item \textbf{There is a strong distance dependence of the X-ray source which severely affects the enclosed mass 
of the collapsing core. Nearby X-ray sources have a smaller negative impact compared to those at larger distances.}
X-ray sources at distances between 1 kpc and 4 kpc all reduce the enclosed mass found
within the core of the collapsing halo compared to the stellar only case. The level of reduction is dependent
on the distance to the source. We found that sources at a distance of 1 kpc suffered approximately a 10 \% 
reduction in enclosed mass while those at distance of 4 kpc suffered a reduction of $\sim 50 \%$. The distance
dependence is a result of the heating effects of the X-rays which results in more diffuse gas and smaller
mass inflow rates. Cold gas which is surrounding the halo when the halo is exposed to only stellar photons is 
heated by the X-rays reducing mass inflow. 
\item \textbf{The \molH formed by the extra free electrons due to X-rays has no material impact on the 
thermodynamics outside the core.} Instead the heating effects of the X-rays are the dominant component. At a 
distance of $\sim 100$ pc from the central density we see more \molH in the X-ray compared to the stellar case 
but the gas is also significantly hotter (see Figure \ref{HaloARayProfile}). The cold gas available for 
accretion in the stellar case has been heated in the X-ray case. This is especially true for sources at an initial 
separation of 2 kpc or greater and hence the larger negative feedback effects in this case.
\item \textbf{Hard X-ray photons from nearby sources can have an additional negative impact.} We found that 
for initial separations of 2 kpc and 4 kpc the inclusion of hard X-rays had a negligible effect on our results and
that their thermal characteristics matches closely that of the soft X-ray models. The source at an initial 
separation of 1 kpc (1kpc\_HX\_A model) resulted in a lower temperature core and a much
lower enclosed mass. This reason for this is that the increased hydrogen ionising ability of the hard X-rays 
in the denser regions of the halo suppresses further the mass inflow rate. The increased electron fraction also 
provides additional \molH causing a slightly lower temperature core, though this effect is small as discussed 
above. Overall, we find that at very close separations hard X-rays have an additional negative feedback effect 
compared to soft X-rays. However, HMXBs accreting at rates comparable to the Eddington rate (say 10\%
Eddington) will produce far more soft X-ray photons than hard X-ray photons. This is because HMXBs which are 
accreting due to Roche lobe overflow will lead to higher disk accretion rates and hence a spectrum peaked at 
lower energies \citep[e.g.][]{Done_2007}. As a result the impact of soft X-ray feedback is likely to be more 
important in the context of DCBH seeds. 
\end{itemize}

\noindent Hence, we conclude that because X-rays do reduce the enclosed mass within the core of the collapsing
halo they can have a negative impact. In particular, and in agreement with previous studies, 
when the source of X-rays is sufficiently distant from the collapsing halo (much like a cosmic X-ray background) 
then there is likely to be a  significant negative feedback effect on forming DCBHs. \textit{However, the caveat 
is that the negative impact diminishes as the distance to the source decreases.} This is an important finding. 
It implies that for close halo pairs \citep{Dijkstra_2008, Agarwal_2012, Dijkstra_2014} or for so-called 
synchronised halo pairs \citep{Visbal_2014b} in an otherwise fairly benign environment the 
negative feedback wrought by X-rays may not be significant due to their close separation. While this may 
further constrain the search for DCBH environments to those regions without a pervasive X-ray background this is 
likely to be the general case at high redshift. Furthermore, the result that sufficiently nearby sources of 
X-rays do not show significant negative feedback further strengthens the case for nearby luminous galaxies 
to be the catalyst for forming DCBHs.

\section*{Acknowledgements}
\noindent J.A.R. would like to thank Chris Done for fruitful and informative discussions on X-ray emission 
mechanisms. This work was supported by the Science and Technology Facilities Council (grant
numbers ST/L00075X/1 and RF040365). This work used the DiRAC Data Centric system at Durham University,  
operated  by  the  Institute  for  Computational  Cosmology on  behalf  of  the  STFC  DiRAC  HPC  
Facility  (www.dirac.ac.uk). This equipment was funded by BIS National E-infrastructure capital grant 
ST/K00042X/1, STFC capital grant ST/H008519/1, and STFC DiRAC Operations grant ST/K003267/1 and 
Durham University.  DiRAC  is  part  of  the  National  E-Infrastructure. 
J.A.R. and P.H.J. acknowledge the support of the Magnus Ehrnrooth Foundation, the Research
Funds of the University of Helsinki and the Academy of Finland grant 1274931.
J.H.W. acknowledges support by NSF and NASA grants AST-1333360,
HST-AR-13895.001, and HST-AR-14326.001. Some of the numerical simulations were also performed on facilities 
hosted by the CSC -IT Center for Science in Espoo, Finland, which are financed by the 
Finnish ministry of education. Computations described in this work were performed using the 
publicly-available \enzo code (http://enzo-project.org), which is the product of a collaborative 
effort of many independent scientists from numerous institutions around the world.  Their 
commitment to open science has helped make this work possible. The freely available astrophysical 
analysis code YT \citep{YT} was used to construct numerous plots within this paper. The authors 
would like to extend their gratitude to Matt Turk et al. for an excellent software package. J.A.R. would like to 
thank Lydia Heck and all of the support staff involved with Durham's COSMA4 and DiRAC's COSMA5 systems for their
technical support. Finally, the authors would like to thank an anonymous referee for a considered and detailed
report which helped to improve both the clarity and quality of this manuscript.

\bsp	% typesetting comment
\label{lastpage}
\end{document}